\documentclass[fleqn,usenatbib]{mnras}
\usepackage{newtxtext,newtxmath}
\usepackage[T1]{fontenc}

\DeclareRobustCommand{\VAN}[3]{#2}
\let\VANthebibliography\thebibliography
\def\thebibliography{\DeclareRobustCommand{\VAN}[3]{##3}\VANthebibliography}

\usepackage{graphicx}	% Including figure files
\usepackage{amsmath}	% Advanced maths commands
\usepackage{amsbsy}
\usepackage{tabularx}
\usepackage{breqn}
\usepackage{mathtools}
\usepackage{multirow}
\usepackage{lineno}
\let\oldequation\equation
\let\oldendequation\endequation
\renewenvironment{equation}
  {\linenomathNonumbers\oldequation}
  {\oldendequation\endlinenomath}

\newcommand\Tstrut{\rule{0pt}{2.9ex}}       % "top" strut
\newcommand\Bstrut{\rule[-1.3ex]{0pt}{0pt}} % "bottom" strut
\newcommand\TBstrut{\Tstrut\Bstrut}         % "top and bottom" strut
\usepackage{xcolor}
\newcommand{\datavector}[0]{\pmb{S}}

\title[Cosmology and neutrino mass with the MST]{Cosmology and neutrino mass with the Minimum Spanning Tree}

\author[K. Naidoo et al.]{
  Krishna Naidoo$^{1,2}$\thanks{E-mail: \href{mailto:knaidoo@cft.edu.pl}{knaidoo@cft.edu.pl}},
  Elena Massara$^{3,4}$, and
  Ofer Lahav$^{2}$\\
  \\
  $^{1}$Center for Theoretical Physics, Polish Academy of Sciences, Al. Lotnik\'{o}w 32/46, 02-668 Warsaw, Poland\\
  $^{2}$Department of Physics \& Astronomy, University College London, Gower Street, London WC1E 6BT, UK\\
  $^{3}$Waterloo Centre for Astrophysics, University of Waterloo, 200 University Ave W, Waterloo, ON N2L 3G1, Canada\\
  $^{4}$Department of Physics and Astronomy, University of Waterloo, Waterloo, ON N2L 3G1, Canada
}

\date{Accepted XXX. Received YYY; in original form ZZZ}

\pubyear{2020}

\begin{document}
\label{firstpage}
\pagerange{\pageref{firstpage}--\pageref{lastpage}}
\maketitle


\begin{abstract}
  The information content of the minimum spanning tree (MST), used to capture higher-order statistics and information from the cosmic web, is compared to that of the power spectrum for a $\nu\Lambda$CDM model. The measurements are made in redshift space using haloes from the Quijote simulation of mass $\geq 3.2\times 10^{13}\,h^{-1}{\rm M}_{\odot}$ in a box of length $L_{\rm box}=1\,h^{-1}{\rm Gpc}$. The power spectrum multipoles (monopole and quadrupole) are computed for Fourier modes in the range $0.006 < k < 0.5\, h{\rm Mpc}^{-1}$. For comparison the MST is measured with a minimum length scale of $l_{\min}\simeq13\,h^{-1}{\rm Mpc}$. Combining the MST and power spectrum allows for many of the individual degeneracies to be broken; on its own the MST provides tighter constraints on the sum of neutrino masses $M_{\nu}$ and cosmological parameters $h$, $n_{\rm s}$, and $\Omega_{\rm b}$ but the power spectrum alone provides tighter constraints on $\Omega_{\rm m}$ and $\sigma_{8}$. Combined we find constraints that are a factor of two (or greater) on all parameters with respect to the power spectrum (for $M_{\nu}$ there is a factor of four improvement). These improvements appear to be driven by the MST's sensitivity to small scale clustering, where the effect of neutrino free-streaming becomes relevant, and high-order statistical information in the cosmic web. The MST is shown to be a powerful tool for cosmology and neutrino mass studies, and therefore could play a pivotal role in ongoing and future galaxy redshift surveys (such as DES, DESI, \emph{Euclid}, and Rubin-LSST).%245 words
\end{abstract}

\begin{keywords}
neutrinos -- cosmological parameters -- large scale structure of Universe
\end{keywords}

%%%%%%%%%%%%%%%%%%%%%%%%%%%%%%%%%%%%%%%%%%%%%%%%%%

%%%%%%%%%%%%%%%%% BODY OF PAPER %%%%%%%%%%%%%%%%%%

%%%%%%%%%%%%%%%%%%%%%%%%%%%%%%%%%%%%%%%%%%%%%%%%%%

\section{Introduction}

The $\Lambda$ Cold Dark Matter ($\Lambda$CDM) paradigm has remained at the forefront of cosmology for over twenty years, cementing it as the standard cosmological model. Observations and simulations over that time have largely strengthened the case for $\Lambda$CDM despite the model consisting overwhelmingly of things we still do not understand -- principally the nature of dark matter and dark energy. While understanding the former will most likely need significant contributions from particle physics experiments, understanding the latter is a key goal for future experiments in cosmology. The next generation of galaxy redshift surveys (such as the Dark Energy Spectroscopic Instrument (DESI),\footnote{\href{http://desi.lbl.gov/}{http://desi.lbl.gov/}} \emph{Euclid},\footnote{\href{http://www.euclid-ec.org/}{http://www.euclid-ec.org/}} the Nancy Grace Roman Space Telescope,\footnote{\href{https://roman.gsfc.nasa.gov}{https://roman.gsfc.nasa.gov}} the Prime Focus Spectrograph,\footnote{\href{https://pfs.ipmu.jp/index.html}{https://pfs.ipmu.jp/index.html}} the Rubin Observatory Legacy Survey of Space and Time,\footnote{\href{https://www.lsst.org/}{https://www.lsst.org/}} and the 4-metre Multi-Object Spectroscopic Telescope\footnote{\href{https://www.4most.eu/cms/}{https://www.4most.eu/cms/}}) will map the positions of hundreds of millions of galaxies. Determining the nature of dark energy is a major scientific mission for these surveys. In $\Lambda$CDM dark energy is simply Einstein's cosmological constant (resulting from a scalar field), but better data may reveal that dark energy is actually changing with time (dynamical dark energy) or that the requirement for dark energy is actually a symptom that General Relativity, on which these models are based, is incomplete and requires modification (the premise of Modified Gravity theories).

The discovery of neutrino oscillation \citep{Kamiokande1998, Sudbury2001} provided evidence that neutrinos are not massless (as had been predicted by the standard model of particle physics). This discovery makes neutrinos of keen interest to particle physicists, as the origins of their mass and hierarchy could provide hints to new physics. Particle physics experiments currently place a lower bound on the sum of neutrino masses (denoted in this work by $M_{\nu}$) of $M_{\nu}\gtrsim 0.06\,{\rm eV}$. However, it is cosmological experiments that provide the tightest upper bound: currently $M_{\nu} \lesssim 0.11\,{\rm eV}$ \citep[95\% confidence level (CL);][]{PlanckPara2018, eBOSS2020}; by contrast the upper bound from particle physics experiment KATRIN is currently $M_{\nu}\leq 1.1\, {\rm eV}$ \citep[90\% CL;][]{Katrin2019} (although unlike cosmological measurements this is model independent). The sensitivity to neutrino mass in cosmology comes from the role neutrinos play in the growth of large scale structure (LSS). The neutrino's characteristic free-streaming length, a quantity dependent on its mass, will wash out small scale structure. This effect can be quantified as a suppression of small scale modes in the power spectrum. Over the next five years surveys such as DESI expect to be sensitive to $M_{\nu} \lesssim 0.06\, {\rm eV}$ \citep[95\% CL;][]{Font2014} and therefore anticipate a first cosmological detection of a non-zero mass for neutrinos.

Cosmological surveys have largely focused on two-point statistics, whether in real or Fourier space. While these methods are tried-and-tested, they fail to fully explore and capture all the information content of galaxy surveys. This is particularly relevant at low redshifts where the highly non-linear structure of the cosmic web is very pronounced; here the distribution of matter cannot be fully characterised by two-point statistics and further statistical methods are required if we are to fully extract all the information present. Such analyses are particularly timely as future data sets will probe the Universe with tracers (galaxies, quasars, etc.) at higher number densities providing a greater sensitivity to the cosmic web. Relevant statistical methods include the three-point correlation function \citep[the Bispectrum in Fourier space; e.g.][]{bispectrum, davideboss2018}, Minkowski functionals \citep[e.g.][]{minkowski}, the 1D probability distribution function \citep{Uhlemann2019}, marked power spectra \citep{Massara2020}, machine learning \citep[ML; e.g.][]{machinelearning}, and the Minimum Spanning Tree \citep[MST;][]{Naidoo2020} -- the focus of this paper.

The MST was first introduced to astronomy by \citet{Barrow1985} and has successfully been used as a filament finder for cosmic web studies \citep{Bhavsar1988, Weygaert1992, Bhavsar1996, Krzewina1996, Ueda1997, Coles1998, Adami1999, Colberg2007, GAMA2014, Beuret2017, Libeskind2018}. The MST is the minimum weighted graph that connects a set of points without forming loops. More recently \citet{Naidoo2020} investigated how the MST could be used to incorporate the cosmic web when constraining cosmological parameters. However, unlike the conventional two-point analysis as performed by most galaxy redshift surveys, reference MST values cannot be calculated analytically and instead need to be calculated from simulations. Fortunately, this problem is not unique to the MST -- conventional statistics such as the power spectrum and bispectrum cannot be computed analytically in the non-linear regime (for Fourier modes $k\gtrsim 0.3\,h{\rm Mpc}^{-1{\tiny }}$) and hence require simulations, as do artificial intelligence (AI) and ML algorithms (as well as other algorithms used to measure non-linear features in the cosmic web). This has created a growing demand for large suites of cosmological simulations and the development of accurate emulators as cosmologists push to extract more information from the distribution of galaxies.

The Quijote simulations \citep{Quijote2019} were designed precisely for this use (i.e. to test new summary statistics such as the MST and AI/ML and to push conventional statistics to smaller scales). In this paper we will use these simulations to measure the information content of the MST. The simulations have previously been used to conduct Fisher matrix analysis for the power spectrum \citep{Quijote2019}, bispectrum \citep{Hahn2019}, 1D probability distribution function \citep{Uhlemann2019}, and marked power spectrum \citep{Massara2020}. In this paper we extend this analysis to the MST; in \citet{Naidoo2020} the MST was tested against measurements of the power spectrum and bispectrum for a few parameters (matter density $\Omega_{\rm m}$, amplitude of scalar fluctuations $A_{\rm s}$, and neutrino mass $M_{\nu}$) to test whether the MST adds new information. Fisher matrix analysis is useful in cosmology as it places a lower bound \citep[the Cramer-Rao bound;][]{Rao1945,Cramer1946} on the uncertainty of cosmological parameters inferred from a given statistic. If the posterior distribution is Gaussian then the Fisher matrix constraints will be realised; otherwise the constraints on any given parameter will be weaker. This analysis is explored for parameters of the $\nu\Lambda$CDM model (i.e. the standard model of cosmology $\Lambda$CDM + massive neutrinos $M_{\nu}$). This will help determine the role that the MST can play in constraining parameters from the current and next generation of galaxy surveys.

The paper is organised as follows. In section \ref{method} we discuss the methodology and data used. In section \ref{result} we present the constraints from components of the MST and demonstrate how including the MST together with measurements of the power spectrum improves parameter constraints in a $\nu\Lambda$CDM model. Finally in section \ref{discussion} we discuss the main results and their implications for cosmology and future surveys.

\section{Method}
\label{method}

In this section we explain the Fisher matrix formalism used to measure the information content of several summary statistics, we explain how we measure the power spectrum multipoles and the MST statistics in redshift space, and we describe properties of the Quijote simulations used in this analysis.

\subsection{Fisher Formalism}

The Fisher matrix \citep{Tegmark1997} $F$ is defined to have elements
\begin{equation}
\label{eq:fisher}
F_{ij} = \sum_{\alpha,\beta} \frac{\partial S_{\alpha}}{\partial \theta_{i}} C_{\alpha\beta}^{-1} \frac{\partial S_{\beta}}{\partial \theta_{j}},
\end{equation}
where $S_{\alpha}$ and $S_{\beta}$ are the elements $\alpha$ and $\beta$ of the data vector $\datavector{}$, $\mathbfss{C}$ is the sample covariance matrix defined to have elements
\begin{equation}
C_{\alpha\beta} = \langle \left(S_{\alpha} - \langle S_{\alpha} \rangle\right) \left(S_{\beta} - \langle S_{\beta} \rangle\right)\rangle,
\label{eq:cov}
\end{equation}
and $\theta_{i}$ and $\theta_{j}$ are parameters $i$ and $j$ of the model. We multiply the inverse of the covariance matrix by the Kaufman-Hartlap factor \citep{Kaufman1967, Hartlap2007} $(N_{\rm sim} - 2 -N_{\rm S})/(N_{\rm sim} - 1)$ where $N_{\rm S}$ is the length of the data vector $\datavector{}$ and $N_{\rm sim}$ is the number of simulations used to estimate the covariance matrix; this compensates for the error in the sample covariance estimation. An implicit assumption of this formalism is that the covariance matrix has no parameter dependence and can be accurately defined by one fiducial point in parameter space.

When reference summary statistics are available analytically, the partial derivatives in the Fisher matrix are straightforward to estimate. However, for some summary statistics (such as the MST) where reference values must be obtained via simulations, the partial derivatives must be estimated numerically; typically we use
\begin{equation}
\label{eq:derivative_eq0}
\frac{\partial \datavector{}}{\partial \theta} \simeq \frac{\datavector{}(\theta+d\theta) - \datavector{}(\theta-d\theta)}{2d\theta} +\mathcal{O}(d\theta^2),
\end{equation}
where $d\theta$ is a small deviation from a fiducial $\theta$. We cannot use this when estimating the partial derivative with respect to neutrino mass when this mass is zero, as this would require simulations with negative neutrino mass; instead here we use one of the estimators
\begin{linenomath} % See https://tex.stackexchange.com/questions/461186
\begin{align}
&\left[\frac{\partial \datavector{}}{\partial M_{\nu}}\right]_{1} \simeq \frac{\datavector{}(dM_{\nu}) - \datavector{}(M_{\nu}=0)}{dM_{\nu}}+\mathcal{O}(dM_{\nu}),\label{eq:derivative_eq1}\\
&\left[\frac{\partial \datavector{}}{\partial M_{\nu}}\right]_{2} \simeq \frac{-\datavector{}(2dM_{\nu}) + 4\datavector{}(dM_{\nu}) - 3\datavector{}(M_{\nu}=0)}{2dM_{\nu}}+\mathcal{O}(dM_{\nu}^{2}),\label{eq:derivative_eq2}\\
&\left[\frac{\partial \datavector{}}{\partial M_{\nu}}\right]_{3} \simeq \frac{\datavector{}(4dM_{\nu}) - 12\datavector{}(2dM_{\nu}) + 32\datavector{}(dM_{\nu}) - 21\datavector{}(M_{\nu}=0)}{12dM_{\nu}}\notag\\
&\quad\quad\quad\quad\quad\quad+\mathcal{O}(dM_{\nu}^{3}).\label{eq:derivative_eq3}
\end{align}
\end{linenomath}
These non-symmetric estimators are designed to use simulations with $M_{\nu}=0.1$, $0.2$, or $0.4\,{\rm eV}$ (in addition to $M_{\nu}=0$). In this case the increment $dM_{\nu}$ can in Eq. \ref{eq:derivative_eq1} be any of these three values, in Eq. \ref{eq:derivative_eq2} can be $M_{\nu}=0.1$ or $0.2\,{\rm eV}$, and in Eq. \ref{eq:derivative_eq3} must be $M_{\nu}=0.1\,{\rm eV}$.

The likelihood is assumed to follow a multivariate Gaussian (e.g. \citet{Heavens2009}, i.e. with Gaussian errors for each parameter) defined by
\begin{equation}
\mathcal{L}(\pmb{\theta}) = \sqrt{\frac{{\rm det} \ \mathbfss{F} }{(2\pi)^M}} \exp \left(-\frac{1}{2}(\pmb{\theta} - \pmb{\theta}_{\rm Fid})^{^{\top}}\cdot\mathbfss{F}\cdot(\pmb{\theta} - \pmb{\theta}_{\rm Fid})\right),
\end{equation}
where $\pmb{\theta}$ (of length $M$) are the parameters of a $\nu\Lambda$CDM model and $\pmb{\theta}_{\rm Fid}$ are fiducial parameters.

\subsection{Summary Statistics in Redshift Space}

In redshift space, redshift space distortions \citep[RSD;][]{Kaiser1987} caused by peculiar velocities alter the observed redshifts of galaxies. This causes a line-of-sight (LOS) shift given by
\begin{equation}
\pmb{x}_{\rm RSD} = \pmb{x} + \frac{1+z}{H(z)}(\pmb{v}\cdot\pmb{e}),
\end{equation}
where $\pmb{x}$ is the real space coordinate, $\pmb{x}_{\rm RSD}$ is the redshift space coordinate, $\pmb{v}$ is the peculiar velocity, $z$ is the redshift, $H(z)$ is the Hubble expansion rate at redshift $z$, and $\pmb{e}$ is the unit vector defining the LOS. In this paper the LOS is taken to be the z-axis ($\pmb{e} = (0, 0, 1)$); while the accuracy and convergence for the partial derivatives estimates is improved by additionally using both the x-axis ($\pmb{e} = (1, 0, 0)$) and y-axis ($\pmb{e} = (0, 1, 0)$) as the LOS \citep{Hahn2019}.

\subsubsection{Power Spectrum Multipoles}

The density field is reexpressed in Fourier space; let $\pmb{k} = (k_{\rm x}, k_{\rm y}, k_{\rm z})$ be a Fourier mode vector. We bin the density field by $k = |\pmb{k}|$ and $\mu = k_{\rm z}/k$ (the cosine of the angle between $\pmb{k}$ and the LOS $\pmb{e}$).

The power spectrum multipoles are
\begin{equation}
P_{\ell}(k) = (2\ell + 1) \int_{0}^{1}P(k, \mu)\mathcal{L}_{\ell}(\mu)d\mu,
\end{equation}
where $\mathcal{L}_{\ell}$ is a Legendre polynomial. The monopole and quadrupole are $P_0$ and $P_2$ respectively.

\subsubsection{Masking Small Scales by Grouping}

Unlike the power spectrum, small scale information cannot be removed by simply cutting the distribution of edge lengths in the MST statistics, instead these scales need to be removed from the input catalogues. To do this, small scales are masked by grouping together points with small separations. This is carried out by grouping two haloes if their separation is less than $l_{\min}=2\pi/k_{\max}=4\pi\,h^{-1}{\rm Mpc}\simeq13\,h^{-1}{\rm Mpc}$ (equivalent to the maximum Fourier mode $k_{\max}=0.5\,h{\rm Mpc^{-1}}$ measured for the power spectrum).
The grouping is transitive (if $A$ and $B$ are close, and $B$ and $C$ are close, then all three are grouped together regardless of the distance between $A$ and $C$); as a result, $l_{\min}$ needs to be set well below the mean separation of points (to avoid the entire catalogue collapsing to one point).
A group of haloes is given coordinates equal to the mean coordinates of its constituent haloes.
This process yields a catalogue of nodes (a collection of grouped and ungrouped haloes); on average this node catalogue is about one quarter the size of the original halo catalogue.
See Appendix \ref{appendix:small_scales} for evidence that this masking technique is effective.

\subsubsection{Minimum Spanning Tree Statistics}

The MST is constructed, in 3D comoving coordinates in redshift space, from the node catalogue. The distribution function $N(x)$ of the MST statistics is measured, where $x$ is the degree $d$, edge length $l$, branch length $b$, or branch shape $s$ \citep{Naidoo2020}. An edge is a line in the MST graph, the degree is the number of edges attached to each node, and a branch is a chain of edges connected continuously by nodes of degree $d=2$. For the branches we measure their length $b$ (i.e. the sum of the lengths of member edges) and their shape $s$ (i.e. the square root of one minus the ratio between the straight line distance between branch ends and the branch length -- with this definition, straighter branches have $s \simeq 0$ while larger values indicate more curved branches). Furthermore, to ensure $N(x)$ can be described by a Gaussian distribution we remove the tails of the distribution function. We measure the mean of the cumulative distribution function (CDF) of $N(x)$ for the fiducial simulations and then measure the MST $N(x)$ only in the region where $0.05<{\rm CDF}<0.95$ (with the exception of $d$ where we include $N(d)$ in the range $1\leq d\leq4$). The publicly available {\sc Python} package {\sc MiSTree} \citep{mistree} was used to construct and measure the statistics of the MST.

\begin{table}
	\centering
	\caption{A summary of the Quijote simulations used in this study, highlighting the deviations from the fiducial cosmological parameters, the type of initial conditions (IC; either first-order perturbation theory -- ZA (Zel'dovich approximation) or second-order perturbation theory -- 2LPT) and the number of realisations.}
	\label{tab:summary_quijote}
	\begin{tabular}{llll}
		\hline
		\TBstrut Name & Deviation from Fiducial & IC & Realisations \\
		\hline
		\Tstrut Fiducial & n/a & 2LPT & 15000\\
    Fiducial_ZA & n/a & ZA & 500\\
		$\Omega_{\rm m}^{+}$ & $\Delta \Omega_{\rm m}=+0.01$ & 2LPT & 500\\
		$\Omega_{\rm m}^{-}$ & $\Delta \Omega_{\rm m}=-0.01$ & 2LPT & 500\\
		$\Omega_{\rm b}^{++}$ & $\Delta \Omega_{\rm b}=+0.002$ & 2LPT & 500\\
		$\Omega_{\rm b}^{--}$ & $\Delta \Omega_{\rm b}=-0.002$ & 2LPT & 500\\
		$h^{+}$ & $\Delta h=+0.02$ & 2LPT & 500\\
		$h^{-}$ & $\Delta h=-0.02$ & 2LPT & 500\\
		$n_{\rm s}^{+}$ & $\Delta n_{\rm s}=+0.02$ & 2LPT & 500\\
		$n_{\rm s}^{-}$ & $\Delta n_{\rm s}=-0.02$ & 2LPT & 500\\
		$\sigma_{\rm 8}^{+}$ & $\Delta \sigma_{\rm 8}=+0.015$ & 2LPT & 500\\
		$\sigma_{\rm 8}^{-}$ & $\Delta \sigma_{\rm 8}=-0.015$ & 2LPT & 500\\
		$M_{\nu}^{+}$ & $\Delta M_{\nu} = +0.1\,{\rm eV}$ & ZA & 500 \\
		$M_{\nu}^{++}$ & $\Delta M_{\nu} = +0.2\,{\rm eV}$ & ZA & 500 \\
		\Bstrut $M_{\nu}^{+++}$ & $\Delta M_{\nu} = +0.4\,{\rm eV}$ & ZA & 500 \\
		\hline
	\end{tabular}
\end{table}

\begin{figure*}
	\centering
	\includegraphics[width=\textwidth]{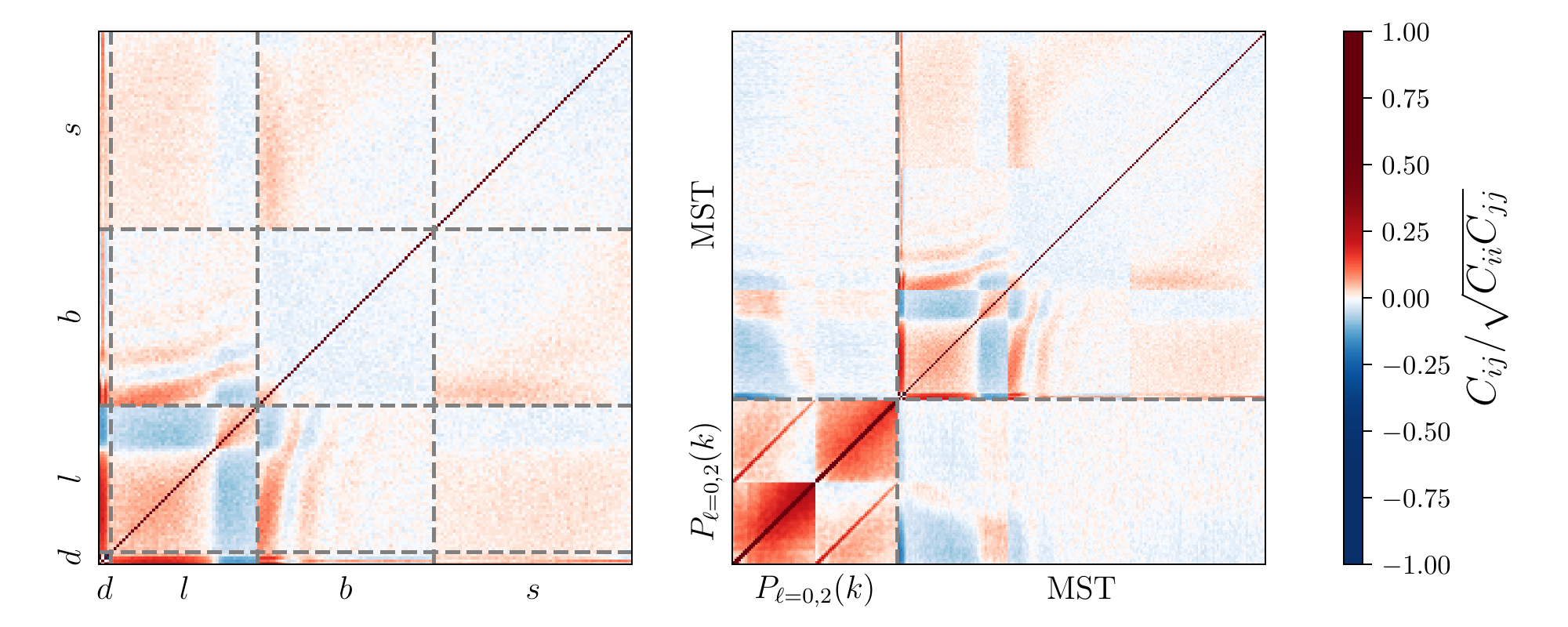}
	\caption{Correlation matrices \textit{(left)} of the MST statistics and \textit{(right)} of the MST statistics and power spectrum multipoles. The components of the MST statistics (the degree $d$, edge length $l$, branch length $b$, and branch shape $s$) are labelled and divided by dashed grey lines on the left. The matrix has significant non-diagonal features. The most striking is the correlation between the edge length $l$ and branch length $b$ where there are several fading lines of positive correlation. These originate from correlations between branches formed from two edges, three edges and so on, becoming fainter for branches formed from more edges. For the degree $d$ and $l$ there is virtually no correlation while there are correlations between $d$ and the two branch statistics length $b$ and shape $s$. These correlations appear to be strongly tied to $d=2$ presumably due to the definition for branches requiring intermediate nodes with degree $d=2$. Lastly there is a weak correlation between $b$ and $s$ and negligible correlations between other MST statistics (i.e. between $l$ and $d$ and between $l$ and $s$). The power spectrum multipoles and components of the MST statistic are divided by a dashed grey lines on the right. The correlation matrix for the power spectrum multipoles is shown to be strongest diagonally, with the most striking feature being the positive correlations between multipoles. Correlations between the power spectrum multipoles and the MST are negligible with only a small inverse correlation seen. This is consistent with expectations since smaller edge lengths should correspond to larger Fourier modes.}
	\label{correlation_matrix}
\end{figure*}

\subsection{Quijote Simulations}

The Quijote simulations \citep{Quijote2019} are a large set of $N$-body simulations designed for quantifying the information content of cosmological observables and for training ML algorithms. The simulations are constructed in boxes of length $L_{\rm box}=1\,h^{-1}{\rm Gpc}$, using $512^{3}$ dark matter particles and $512^{3}$ neutrino particles (for simulations with massive neutrinos). A detailed table of the parameters used for the Quijote simulations can be found in Tab. 1 of \citet{Quijote2019}. The simulations are based on a fiducial $\Lambda$CDM cosmology (based on \citealt{PlanckPara2018}) with matter density $\Omega_{\rm m}=0.3175$, baryon density $\Omega_{\rm b}=0.049$, Hubble constant $h=0.6711$, primordial spectral tilt $n_{\rm s}=0.9624$, the root mean square of the linear power spectrum at spheres of radius $8\,h^{-1}{\rm Mpc}$ $\sigma_{8}=0.834$, sum of neutrino masses $M_{\nu}=0\,{\rm eV}$, and dark energy equation of state $w=-1$. The power spectrum multipoles and MST are computed on haloes with masses larger than $3.2\times 10^{13}\,h^{-1}{\rm M}_{\odot}$. For each parameter we determine the dependence with respect to that parameter using 500 simulations in which only that parameter deviates from its fiducial value (while maintaining zero curvature, i.e. a shift in $\Omega_{\rm m}\implies\Omega_{\rm \Lambda}=1-\Omega_{\rm m}$).
To construct the covariance matrix we use 15,000 \textit{fiducial simulations} constructed with the fiducial cosmology. See Tab. \ref{tab:summary_quijote} for a summary. Simulations with massive neutrinos ($M_{\nu}^{+}$, $M_{\nu}^{++}$, and $M_{\nu}^{+++}$) are produced from simulations with initial conditions following first-order perturbation theory (i.e. the Zel'dovich approximation -- ZA) instead of second-order Lagrangian perturbation theory (2LPT) since the 2LPT approach is not implemented for massive neutrino cosmologies \citep[for further details see][]{Quijote2019}. Therefore to remove any potential systematic bias this discrepancy may present we will be using the Fiducial_ZA simulations for the neutrino partial derivative estimates, rather than the Fiducial simulations, since these simulations are also produced with ZA initial conditions.

\section{Results}
\label{result}

This Section discusses the following results: (1) the covariance matrix for the MST statistics and the internal correlations and correlations with the power spectrum, (2) the partial derivatives of the power spectrum and MST statistics, and (3) parameter constraints for a $\nu\Lambda$CDM model obtained from individual and combined measurements of the MST and power spectrum.

\subsection{Covariance Matrix}
\label{covariance_matrix}

The covariance matrix is constructed from Eq. \ref{eq:cov} using data vectors measured from 15,000 fiducial simulations. Fig. \ref{correlation_matrix} shows the correlation matrix for the MST on the left and the correlation matrix for the combined data vector of the power spectrum multipoles and MST statistics on the right. Unlike the correlation matrix for the power spectrum, the correlation matrix for the MST contains several non-diagonal features. One of the most striking features is the correlation between the edge length $l$ and branch length $b$ which show `waves' of positive correlations between short edges and short branches followed by negative correlations and then positive correlation for longer edges and branches. These positive correlations stem from the correlations between branches formed from two edges, three edges, and so on. For branches formed from more edges these correlations become weaker as branches formed from more than three edges are rare. Other correlations in the MST statistics appear to stem from branches, which by definition have intermediate nodes with degree $d=2$. As a result we see strong correlations between the degree and branch length. The correlation between branch length and shape is weak but indicates that longer branches are more curved than short ones. The correlations between the power spectrum multipoles and MST are weak so adding the MST to $P(k)$ is beneficial. This is clearest for the monopole and edge lengths which show an inverse correlation between edges and Fourier modes; this is completely consistent with the inverse relation between Fourier space and real space. Furthermore the large scale modes of the monopole (the first half of the data vector) show positive correlations with longer edges. This indicates that most of the large scale clustering information is stored in the large edges of the MST.

\begin{figure*}
	\centering
	\includegraphics[width=\textwidth]{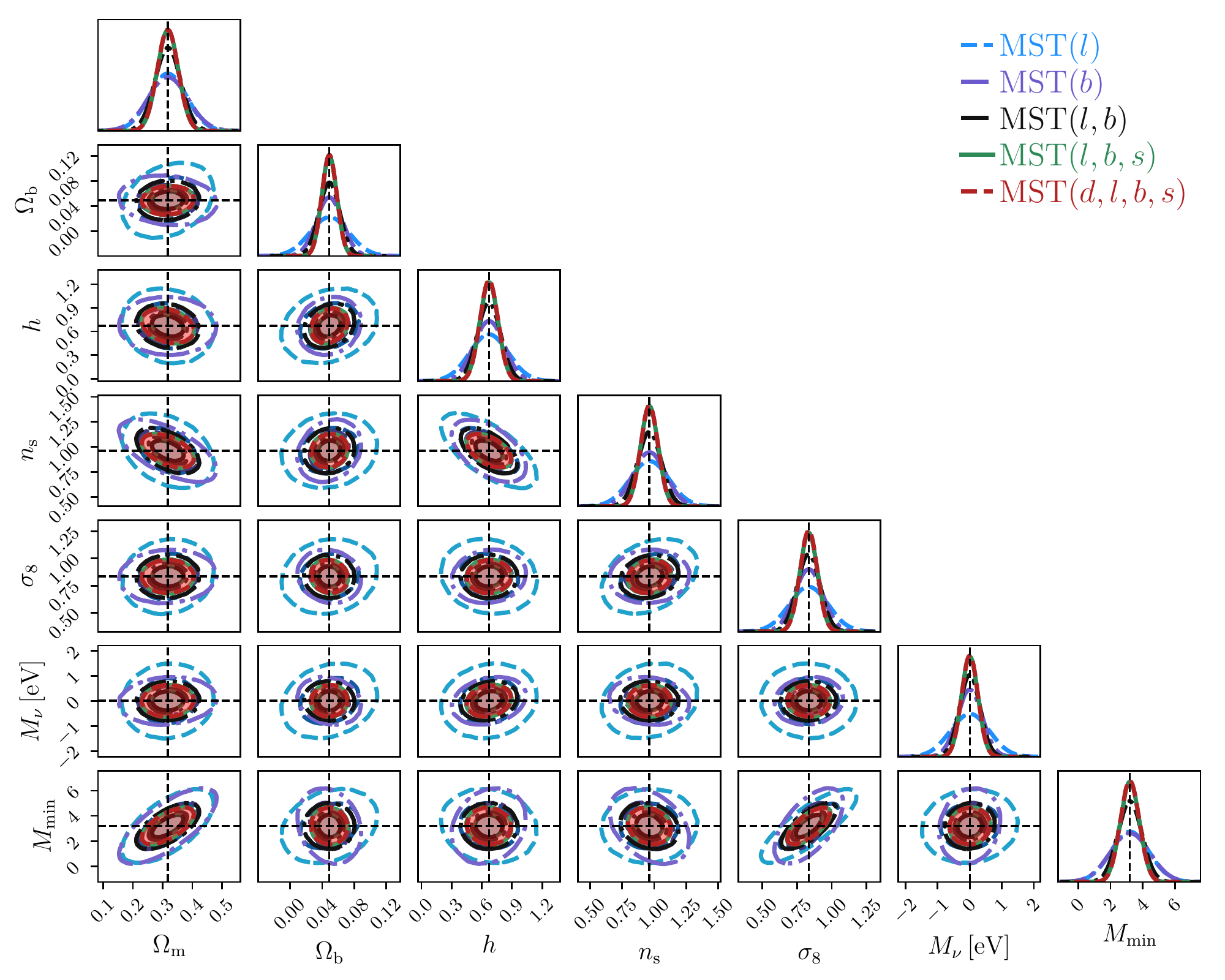}
	\caption{Fisher matrix constraints for $\nu\Lambda$CDM parameters from the MST edge and branch length distributions in combination with the distribution of degree and branch shape in redshift space at $z=0.5$. The individual constraints from the edge ${\rm MST}(l)$ and branch length distributions ${\rm MST}(b)$ are shown in blue and purple respectively. The combined constraint from the edge and branch lengths ${\rm MST}(l, b)$ is shown with black dashed lines, from the edge, branch length and degree distributions ${\rm MST}(l, b, s)$ with green lines, and from the edge, branch length, degree and branch shape distribution ${\rm MST}(d, l, b, s)$ with red lines and contours. The contours show that adding the degree does very little to improve the constraints from the edge and branch length distribution while the branch shape tightens constraints overall.}
	\label{mst_dlbs_RSD}
\end{figure*}

\begin{figure*}
	\centering
	\includegraphics[width=\textwidth]{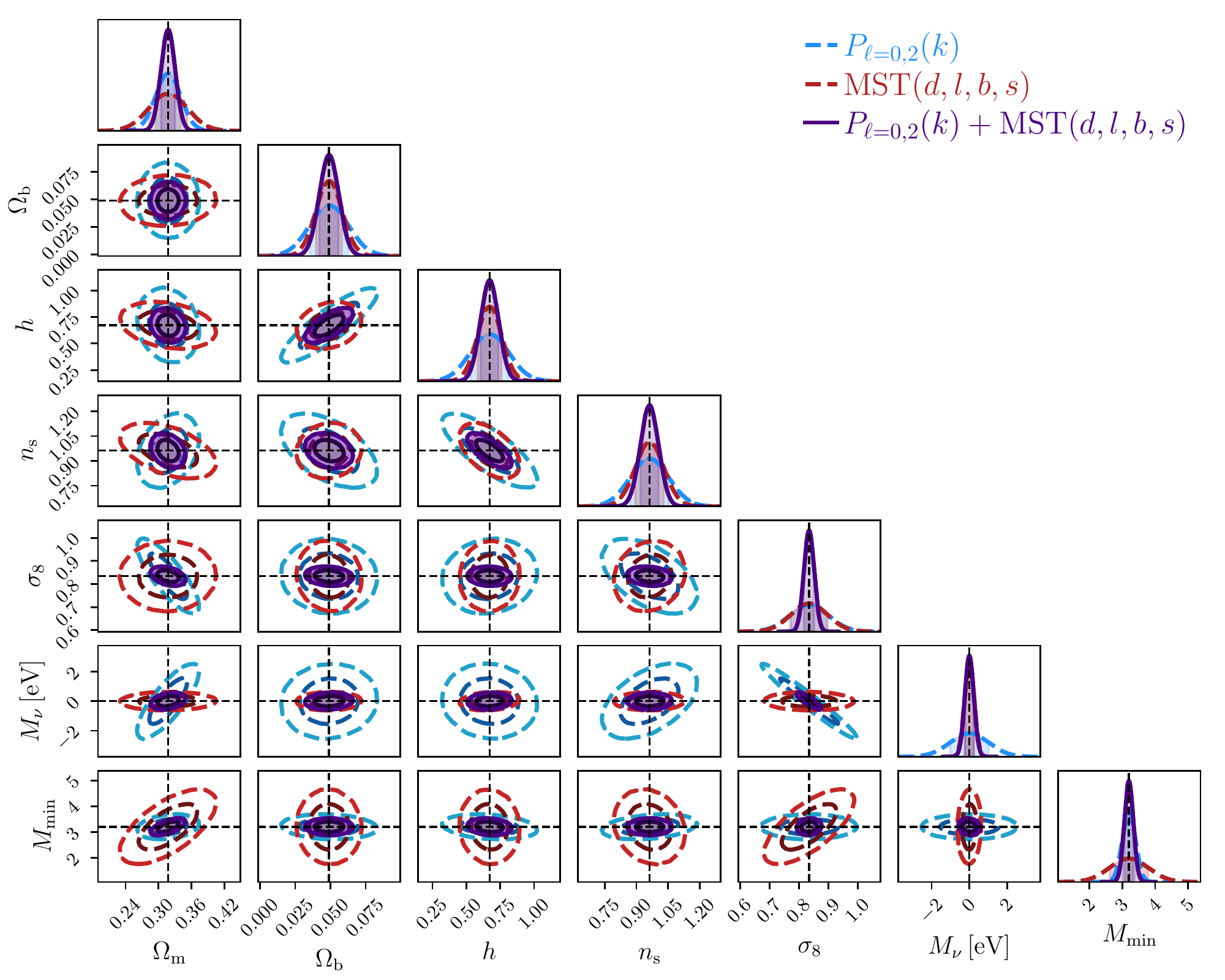}
	\caption{Fisher matrix constraints on $\nu\Lambda$CDM parameters from the power spectrum and MST separately and combined in redshift space at $z=0.5$. The constraints from the power spectrum multipoles $P_{\ell=0,2}(k)$ are shown with blue dashed lines, from the four MST statistics ${\rm MST}(d, l, b, s)$ with red dotted lines, and from the combination of the power spectrum multipoles and MST $P_{\ell=0,2}(k)+{\rm MST}(d, l, b, s)$ with purple lines and contours. Constraints for $\sigma_{8}$ and $\Omega_{\rm m}$ are dominated by the power spectrum multipoles while $h$, $n_{\rm s}$, $\Omega_{\rm b}$, and $M_{\nu}$ are dominated by the MST. Significant degeneracies are broken when combined, leading to much tighter constraints (in comparison to the individual constraints from the power spectrum multipoles) on $h$, $n_{\rm s}$, $\Omega_{\rm m}$, and $\Omega_{\rm b}$ (which improve by a factor of $\sim 2$); $\sigma_{8}$ (which improves by a factor of $\sim 3.89$) and $M_{\nu}$ (which improves by a factor of $\sim 4.35$).}
	\label{pk_mst_RSD}
\end{figure*}

\begin{figure*}
	\centering
	\includegraphics[width=\textwidth]{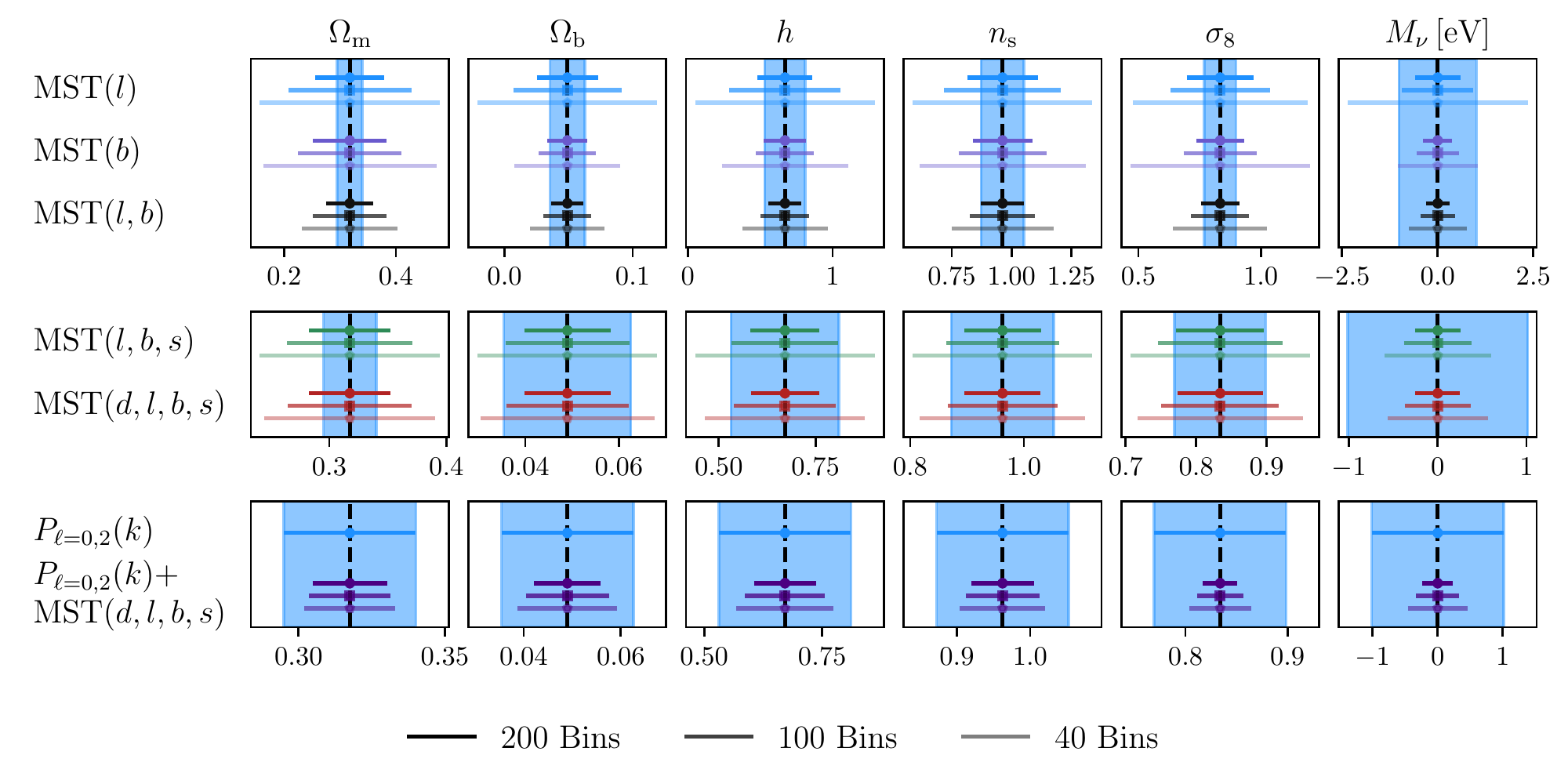}
	\caption{The one dimensional parameter constraints from the MST statistics individually and in combination with the power spectrum multipoles (measured at redshift $z=0.5$). From left to right we show the constraints on the cosmological parameters of the $\nu\Lambda$CDM model: $\Omega_{\rm m}$, $\Omega_{\rm b}$, $h$, $n_{\rm s}$, $\sigma_{8}$, and $M_{\nu}$. From top to bottom we show the constraints for the individual MST statistics: edge length ${\rm MST}(l)$ and branch length ${\rm MST}(b)$, then constraints from the combined MST statistics: edge and branch length ${\rm MST}(l, b)$, with the addition of branch shape ${\rm MST}(l, b, s)$ and with the addition of the degree ${\rm MST}(d, l, b, s)$, and finally in combination with the power spectrum: first showing the power spectrum multipoles alone $P_{\ell=0,2}(k)$ (these are marked with blue envelopes for comparison to the other statistics) and in combination $P_{\ell=0,2}(k)+{\rm MST}(d,l,b,s)$. These are further subdivided for constraints involving the MST with different binning schemes $N_{\rm bins}$: 200 bins is shown with a dark coloured error bar at the top, 100 bins is shown with slightly paler coloured error bars in the middle and 40 bins is shown with the palest coloured error bars at the bottom. These results indicate the significance that binning will have on the individual MST constraints; they also show that, regardless of binning, combining MST statistics with the power spectrum significantly improves constraints on all parameters.}
	\label{constraint_summary_plot}
\end{figure*}

\subsection{Fisher Matrix and Partial Derivatives}

The Fisher matrix is calculated from Eq. \ref{eq:fisher} from the data vector $\datavector{}$. In this Section the data vectors are either the power spectrum monopole and quadrupole $P_{\ell=0,2}(k)$ or combinations of the MST statistics: degree $d$, edge length $l$, branch length $b$, and branch shape $s$. To calculate the Fisher matrix we require measurements of the derivatives $\partial\datavector{}/\partial\theta$; these are estimated using simulations summarised in Tab. \ref{tab:summary_quijote}. For each set of simulations (consisting of 500 individual simulations and 15,000 simulations for the fiducial set) the mean and standard deviation of the summary statistics are obtained. The derivatives for the parameters are then obtained using Eq. \ref{eq:derivative_eq0}, with the exception of neutrino mass where three estimators are used (Eqs. \ref{eq:derivative_eq1}, \ref{eq:derivative_eq2}, and \ref{eq:derivative_eq3}). In real data we would not be able to place a clean cut on the galaxy masses; instead, these limits would be imposed by survey designs and magnitude limits. To account for this in our analysis we follow \citet{Hahn2019} by adding the nuisance parameter $M_{\min}$ which characterises the dependence on the minimum halo mass which is calculated by running the statistics on the fiducial suite with $M_{\min} = 3.1$ and $M_{\min}=3.3$ $\times 10^{13}h^{-1}{\rm M_{\odot}}$. However, unlike in the study by \cite{Hahn2019} we do not include a linear bias nuisance parameter (as it is not clear how a linear bias would effect the MST statistics and including it solely in the power spectrum measurements would weaken the constraints in the power spectrum but leave the MST unaffected, in effect biasing our results in favour of the MST). To avoid this we use only the minimum halo mass as a nuisance parameter in this study.

The more accurate estimators for the partial derivatives are those for which the errors are given to higher orders of $d\theta$. For most parameters the derivative $\partial\datavector{}/\partial\theta$ is determined by the symmetric derivative estimator Eq. \ref{eq:derivative_eq0} which has errors of order $\mathcal{O}(d\theta^{2})$, while the most accurate estimator for $\partial\datavector{}/\partial M_{\nu}$ is given by Eq. \ref{eq:derivative_eq3} which has errors of order $\mathcal{O}(d\theta^{3})$. However, in order for the estimators to be consistent for all parameters the appropriate estimator to use for $M_{\nu}$ is Eq. \ref{eq:derivative_eq2} which has errors of order  $\mathcal{O}(d\theta^{2})$. Previous studies \citep{Quijote2019, Uhlemann2019, Hahn2019} have used Eq. \ref{eq:derivative_eq3} (while \citealt{Massara2020} uses Eq. \ref{eq:derivative_eq2}), so to facilitate comparisons to these studies the results obtained using this estimator are additionally provided in Tab. \ref{tab:summarise_constraints}. However, throughout this paper we will refer to results obtained using Eq. \ref{eq:derivative_eq2} with $dM_{\nu}=0.2\,{\rm eV}$. The partial derivatives for the data vectors (as a function of the $\nu\Lambda$CDM parameters $\Omega_{\rm m}$, $\Omega_{\rm b}$, $h$, $n_{\rm s}$,  $\sigma_{8}$, and $M_{\nu}$) are shown in appendix \ref{appendix:partial_derivatives}.

We test for convergence of the partial derivatives by estimating the Fisher matrix from a fraction of the total simulations available. For the power spectrum we find that partial derivatives are converged for all $M_{\nu}$ derivative estimates and for all parameters while for the MST we find that Eq. \ref{eq:derivative_eq2} with $dM_{\nu}=0.2\,{\rm eV}$ provides the most reliable converged estimates at all redshifts (see Fig. \ref{mst_derivative_convergence}) and quote results from this estimator unless stated otherwise.

\subsection{MST Bin Size Dependence}

We test the sensitivity of the MST distribution functions to the bin size. In real data the size of bins would be dictated by numerical constraints, i.e. the number of mocks used to obtain the covariance matrices and the sample size. Since the suite used to estimate the covariance matrix is large we do not need to keep the data vector short and therefore can test the sensitivity to the MST distribution function's bin size. To this end we use three binning strategies for the distribution functions of $l$, $b$ and $s$, with $N_{\rm bins}=200$, $100$, and $40$ bins. The constraints from the different binning strategies are described and discussed in the following sections (in particular see Fig. \ref{constraint_summary_plot}).

\subsection{Sensitivity to Neutrinos and {$\Lambda$}CDM}
\label{fisher_results}

In this Section the forecast constraints (derived from Fisher matrices) for parameters of a $\nu\Lambda$CDM universe are obtained and discussed.

\subsubsection{Constraints from the Minimum Spanning Tree}

In Fig. \ref{mst_dlbs_RSD} the individual and combined constraints from the edge length $l$, branch length $b$, branch shape $s$, and degree $d$ distributions are shown using $l$, $b$, and $s$ with $N_{\rm bins}=200$. The edge and branch length distributions have constraints that are competitive with each other, with branch length providing but that in combination are significantly tighter. In augmenting the constraints first with the branch shape and then with the degree we see that much of the constraining power for the MST is contained in the edge and branch length distribution; the addition of the branch shape adds a modest improvement to the overall constraints ($\sim 25\%$ for cosmological parameters at $z=0.5$) but the addition of the degree appears to provide a negligible improvement ($\sim 1\%$ for cosmological parameters at $z=0.5$).

\subsubsection{Combined Constraints from the Minimum Spanning Tree and Power Spectrum}

The separate and combined constraints from the MST (using $l$, $b$, and $s$ with $N_{\rm bins}=200$) and power spectrum measured at redshift $z=0.5$ are shown in Fig. \ref{pk_mst_RSD} (for the constraints obtained from different redshifts refer to Tab. \ref{tab:summarise_constraints}). The constraints obtained for $\Omega_{\rm b}$, $h$, $n_{\rm s}$, and neutrino mass $M_{\nu}$ are significantly tighter for the MST, while the power spectrum yields better constraints on $\Omega_{\rm m}$ and $\sigma_{8}$. When the two are combined, significant degeneracies are broken leading to much tighter constraints than those obtained using either statistic alone. For neutrino mass $M_{\nu}$ the combination provides a $1\sigma$ constraint that is $4.35\times$ tighter, for $\Omega_{\rm m}$ that is $1.77\times$ tighter, for $\Omega_{\rm b}$ that is $2.06\times$ tighter, for $h$ that $2.12\times$ tighter, for $n_{\rm s}$ that is $2.09 \times$ tighter, and for $\sigma_{8}$ that is $3.82\times$ tighter.

The impact of the MST binning is explored in Fig. \ref{constraint_summary_plot}, where we display the marginalized constraints for the MST statistics (individually and for different combinations) at redshift $z=0.5$ in comparison to the power spectrum multipoles with different levels of binning. Larger bins are associated with poorer constraints but in each case combining with the power spectrum still leads to significant improvements in the constraints of all parameters. This highlights the importance of including the MST in future galaxy redshift surveys to test and constrain $\Lambda$CDM parameters and to determine the sum of neutrino mass $M_{\nu}$ -- a key scientific goal of many future surveys.

\begin{figure}
	\centering
	\includegraphics[width=\columnwidth]{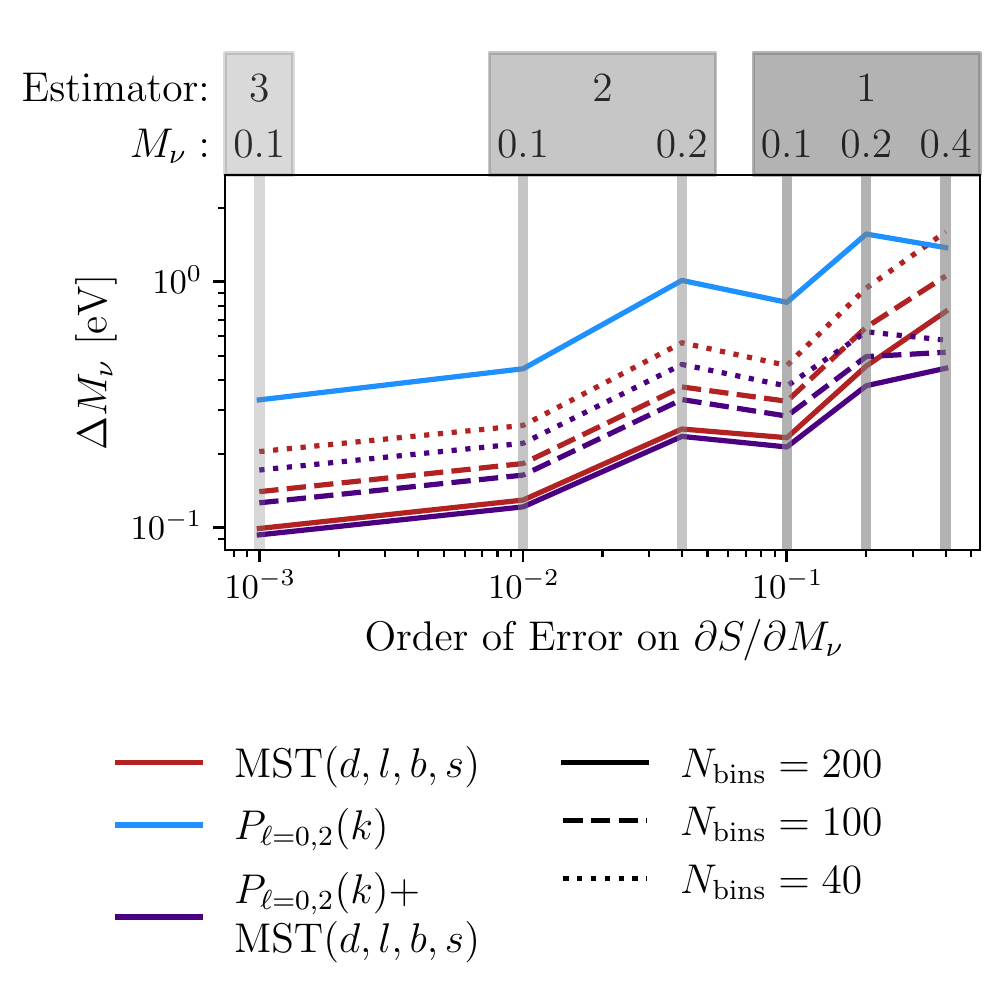}
	\caption{The constraints on neutrino mass from different partial derivative estimates for the MST and power spectrum measured at redshift $z=0.5$ (individually and in combination). The MST statistics are shown with three binning schemes ($N_{\rm bins}=$ 200, 100 and 40). The constraints are given as a function of the order of the absolute error on $\partial\datavector{}/\partial M_{\nu}$. Estimator 1 is given by Eq. \ref{eq:derivative_eq1}, Estimator 2 by Eq. \ref{eq:derivative_eq2}, and Estimator 3 by Eq. \ref{eq:derivative_eq3}. The estimators are indicated by shaded grey lines with the value of $M_{\nu}$ used for each estimator labelled. With the exception of the 40 bins data, the MST consistently outperforms the power spectrum constraints. However, irrespective of the bin size, significant improvements in the constraints of neutrino mass are obtained when combining the MST with the power spectrum.}
	\label{fig_mnu_derivative_comparison}
\end{figure}

\subsubsection{Sensitivity to the Neutrino Partial Derivative Estimator}

The partial derivative of the summary statistics as a function of neutrino mass can be estimated using Eqs. \ref{eq:derivative_eq1}, \ref{eq:derivative_eq2}, and \ref{eq:derivative_eq3}. Fig. \ref{fig_mnu_derivative_comparison} compares the neutrino mass constraints $\Delta M_{\nu}=\sqrt{F_{\nu\nu}^{-1}}$ using the different estimators for the power spectrum and MST measured at redshift $z=0.5$ (see Tab. \ref{mv_estimator_constraints} for a full summary of the different estimators for different components and combinations of the MST statistics at different redshifts). The intrinsic accuracy of the estimators appear to show that less accurate estimators are associated with poorer constraints. The most accurate estimators, i.e. estimator 2 and 3 with $M_{\nu}=0.1\,{\rm eV}$, are shown to be fairly consistent, with estimator 3 generally providing tighter constraints. Without prior knowledge of the true partial derivative it is difficult to know which estimator better captures the true behaviour. For consistency we focus on the results from estimator 2 (with $M_{\nu}=0.2\,{\rm eV}$) which has the same absolute errors as the estimator used on the other parameters (Eqs. \ref{eq:derivative_eq0}) and has been found to have the best convergence properties at all redshifts (see Appendix \ref{appendix:partial_derivatives}). Nevertheless as is shown in Fig. \ref{fig_mnu_derivative_comparison} the relative strengths of the constraints from the power spectrum and MST individually and combined remains, meaning our overall conclusions are insensitive to the neutrino derivative estimator used.

In all cases the MST constraints on $M_{\nu}$ are stronger than ones obtained from the power spectrum. If more bins are used for the MST then the combined constraints with the power spectrum are significantly improved in comparison to the power spectrum alone.

\section{Conclusion}
\label{discussion}

Using haloes from the Quijote simulations we calculate the information content (from the Fisher matrix) of the MST statistics in a $\nu\Lambda$CDM model. The Quijote simulations are a large suite of $N$-body simulations designed to test the information content of summary statistics and to test AI/ML algorithms. In this paper we use a subset of the full suite of simulations (summarised in Tab. \ref{tab:summary_quijote}), measuring the power spectrum multipoles and MST statistics in redshift space on haloes with mass $\geq 3.1\times 10^{13}\,h^{-1}{\rm M}_{\odot}$. The analysis presented in this paper is based on the Fisher matrix formalism which considers only second derivatives of the data vector, rather than a full exploration of the parameter space via for e.g. MCMC. Fisher matrices are limited in the sense that they place only lower bounds (the Cramer-Rao bound) on the uncertainties of cosmological parameters and assume the covariance matrix has no cosmological parameter dependence; as such analysis on real data is likely to depart significantly due to non-Gaussianities \citep[see][]{Hawken2012, Foroozan2021}.

From the MST we measure the distribution of degree $d$, edge length $l$, branch length $b$, and branch shape $s$. In \citet{Naidoo2020} it was shown that $l$ is the most constraining MST statistic. However, this previous analysis was based on COLA simulations of length $L_{\rm box}=250\,h^{-1}{\rm Mpc}$ and limited to the most massive 5000 haloes. Due to the size of the sample, the study saw only modest improvements of 17\% on the $1\sigma$ constraints on $\Omega_{\rm m}$ and 12\% on the $1\sigma$ constraints on $A_{\rm s}$. By using the Quijote simulations we are able to expand this analysis to a wider set of cosmological parameters ($\Omega_{\rm m}$, $\Omega_{\rm b}$, $h$, $n_{\rm s}$, $\sigma_{8}$, and $M_{\nu}$ compared to only $A_{\rm s}$, $\Omega_{\rm m}$, and $M_{\nu}$), over a larger volume ($L_{\rm box}=1\,h^{-1}{\rm Gpc}$), and using a larger catalogue of haloes (of the order of $10^{5}$). In Fig. \ref{mst_dlbs_RSD} the constraints from the MST are shown to be dominated by the distribution of edge and branch lengths, with the degree adding very little information and the branch shape providing a moderate improvement overall.

In Fig. \ref{pk_mst_RSD} and Tab. \ref{tab:summarise_constraints} the MST and power spectrum constraints are compared and combined. The power spectrum multipoles provide much stronger constraints on $\Omega_{\rm m}$ and $\sigma_{8}$. On the other hand the MST dominates the constraints on $M_{\nu}$: at $z=0.5$ the power spectrum multipoles gives $\Delta M_{\nu}=1\,{\rm eV}$ ($1\sigma$), the MST gives $\Delta M_{\nu}=0.25\,{\rm eV}$ ($1\sigma$), and the power spectrum multipoles and the MST combined give $\Delta M_{\nu}=0.23\,{\rm eV}$ -- an improvement of a factor of $\sim4.35$. For the other parameters we find that combining the two sets of statistics breaks several degeneracies and thereby improves constraints on $h$, $n_{\rm s}$, $\Omega_{\rm m}$ and $\Omega_{\rm b}$ by a factor of $\sim 2$ and $\sigma_{8}$ by a factor of $\sim 3.82$ in comparison to the individual power spectrum constraints. We measure the dependence on the number of bins $N_{\rm bins}$, showing that fewer bins will decrease the constraining power but that the inclusion of the MST will, irrespective of the bin number, significantly improve constraints on all parameters.

In this paper we show that adding the MST to the power spectrum greatly improves constraints for parameters of the $\nu\Lambda$CDM model. In particular constraints on $M_{\nu}$ are dominated by the MST since it is significantly more sensitive to the effects of neutrinos on the distribution of haloes.
This appears to come from the sensitivity of the MST to extra information contained in the higher-order statistics of the cosmic web. \citet{Bonnaire2021} have shown that filaments in the cosmic web are the most sensitive environment to neutrino mass. Given the MST propensity for filament detection, we believe the sensitivity to the neutrino mass is related to the detection of filamentary structures by the MST. However, only future work comparing both cosmic web environments from the density field and MST structures from halos will be able to determine this for certain.

Future galaxy surveys such as DESI, \emph{Euclid}, and LSST are projected to bring the upper limit on neutrino mass below or close to $0.06\,{\rm eV}$ \citep{Font2014}, the lower limit from neutrino oscillations (assuming normal hierarchy). As we have illustrated using halo simulations the MST can improve by a factor of four the constraints on neutrino mass with respect to just the power spectrum. Therefore we can expect the MST to provide greater constraints from current and future surveys, possibly enabling this interesting regime to be probed sooner. These constraints could be further improved by combining the MST and the power spectrum with other probes such as the CMB.
This demonstrates the importance of measuring more than just the power spectrum (or two-point statistics) and provides a powerful argument for making measurements of the MST on current and future galaxy surveys such as BOSS, eBOSS, and DESI. In future work we will look to develop techniques for addressing the challenges associated with real galaxy survey data; these challenges include the impact of the survey's selection function across the sky (completeness and depth) and as a function of redshift, the incorporation of galaxy weights, the mitigation of small scale effects including the effect of fibre collisions in spectroscopic surveys, the modeling of galaxy bias and halo occupation distribution parameters, and the development of MST emulators or likelihood free techniques for computing the posterior probability for real data.

\section*{ACKNOWLEDGEMENTS}

We thank Lorne Whiteway for his generous contributions to the development of this paper and for providing insightful suggestions and comments. We thank Peter Coles and Andrew Pontzen for providing useful comments and discussions. KN acknowledges support from the Science and Technology Facilities Council grant ST/N50449X and from the (Polish) National Science Centre grant \#2018/31/G/ST9/03388. OL acknowledges support from an STFC Consolidated Grant ST/R000476/1.

\section*{Data Availability}

In this paper we use halo simulation catalogues from the Quijote simulation. Information on how to access these data products can be found on the Quijote documentation page\footnote{\href{https://quijote-simulations.readthedocs.io/en/latest/index.html}{https://quijote-simulations.readthedocs.io/en/latest/index.html}}.

%%%%%%%%%%%%%%%%%%%%% REFERENCES %%%%%%%%%%%%%%%%%%
%
%% The best way to enter references is to use BibTeX:
%
\bibliographystyle{mnras}
\bibliography{bibfile} % if your bibtex file is called example.bib

\begin{thebibliography}{}
\makeatletter
\relax
\def\mn@urlcharsother{\let\do\@makeother \do\$\do\&\do\#\do\^\do\_\do\%\do\~}
\def\mn@doi{\begingroup\mn@urlcharsother \@ifnextchar [ {\mn@doi@}
  {\mn@doi@[]}}
\def\mn@doi@[#1]#2{\def\@tempa{#1}\ifx\@tempa\@empty \href
  {http://dx.doi.org/#2} {doi:#2}\else \href {http://dx.doi.org/#2} {#1}\fi
  \endgroup}
\def\mn@eprint#1#2{\mn@eprint@#1:#2::\@nil}
\def\mn@eprint@arXiv#1{\href {http://arxiv.org/abs/#1} {{\tt arXiv:#1}}}
\def\mn@eprint@dblp#1{\href {http://dblp.uni-trier.de/rec/bibtex/#1.xml}
  {dblp:#1}}
\def\mn@eprint@#1:#2:#3:#4\@nil{\def\@tempa {#1}\def\@tempb {#2}\def\@tempc
  {#3}\ifx \@tempc \@empty \let \@tempc \@tempb \let \@tempb \@tempa \fi \ifx
  \@tempb \@empty \def\@tempb {arXiv}\fi \@ifundefined
  {mn@eprint@\@tempb}{\@tempb:\@tempc}{\expandafter \expandafter \csname
  mn@eprint@\@tempb\endcsname \expandafter{\@tempc}}}

\bibitem[\protect\citeauthoryear{{Adami} \& {Mazure}}{{Adami} \&
  {Mazure}}{1999}]{Adami1999}
{Adami} C.,  {Mazure} A.,  1999, \mn@doi [\aaps] {10.1051/aas:1999145}, \href
  {https://ui.adsabs.harvard.edu/abs/1999A&AS..134..393A} {134, 393}

\bibitem[\protect\citeauthoryear{{Ahmad} et~al.,}{{Ahmad}
  et~al.}{2001}]{Sudbury2001}
{Ahmad} Q.~R.,  et~al., 2001, Physical Review Letters, \href
  {http://adsabs.harvard.edu/abs/2001PhRvL..87g1301A} {87, 071301}

\bibitem[\protect\citeauthoryear{{Aker} et~al.,}{{Aker}
  et~al.}{2019}]{Katrin2019}
{Aker} M.,  et~al., 2019, \mn@doi [\prl] {10.1103/PhysRevLett.123.221802},
  \href {https://ui.adsabs.harvard.edu/abs/2019PhRvL.123v1802A} {123, 221802}

\bibitem[\protect\citeauthoryear{{Alam} et~al.,}{{Alam}
  et~al.}{2021}]{eBOSS2020}
{Alam} S.,  et~al., 2021, \mn@doi [\prd] {10.1103/PhysRevD.103.083533}, \href
  {https://ui.adsabs.harvard.edu/abs/2021PhRvD.103h3533A} {103, 083533}

\bibitem[\protect\citeauthoryear{{Alpaslan} et~al.,}{{Alpaslan}
  et~al.}{2014}]{GAMA2014}
{Alpaslan} M.,  et~al., 2014, \mn@doi [\mnras] {10.1093/mnras/stt2136}, \href
  {http://adsabs.harvard.edu/abs/2014MNRAS.438..177A} {438, 177}

\bibitem[\protect\citeauthoryear{{Barrow}, {Bhavsar}  \& {Sonoda}}{{Barrow}
  et~al.}{1985}]{Barrow1985}
{Barrow} J.~D.,  {Bhavsar} S.~P.,   {Sonoda} D.~H.,  1985, \mn@doi [\mnras]
  {10.1093/mnras/216.1.17}, \href
  {https://ui.adsabs.harvard.edu/abs/1985MNRAS.216...17B} {216, 17}

\bibitem[\protect\citeauthoryear{{Beuret}, {Billot}, {Cambr{\'e}sy}, {Eden},
  {Elia}, {Molinari}, {Pezzuto}  \& {Schisano}}{{Beuret}
  et~al.}{2017}]{Beuret2017}
{Beuret} M.,  {Billot} N.,  {Cambr{\'e}sy} L.,  {Eden} D.~J.,  {Elia} D.,
  {Molinari} S.,  {Pezzuto} S.,   {Schisano} E.,  2017, \mn@doi [\aap]
  {10.1051/0004-6361/201629199}, \href
  {https://ui.adsabs.harvard.edu/abs/2017A&A...597A.114B} {597, A114}

\bibitem[\protect\citeauthoryear{{Bhavsar} \& {Ling}}{{Bhavsar} \&
  {Ling}}{1988}]{Bhavsar1988}
{Bhavsar} S.~P.,  {Ling} E.~N.,  1988, \mn@doi [\pasp] {10.1086/132325}, \href
  {http://adsabs.harvard.edu/abs/1988PASP..100.1314B} {100, 1314}

\bibitem[\protect\citeauthoryear{{Bhavsar} \& {Splinter}}{{Bhavsar} \&
  {Splinter}}{1996}]{Bhavsar1996}
{Bhavsar} S.~P.,  {Splinter} R.~J.,  1996, \mn@doi [\mnras]
  {10.1093/mnras/282.4.1461}, \href
  {http://adsabs.harvard.edu/abs/1996MNRAS.282.1461B} {282, 1461}

\bibitem[\protect\citeauthoryear{{Bonnaire}, {Aghanim}, {Kuruvilla}  \&
  {Decelle}}{{Bonnaire} et~al.}{2021}]{Bonnaire2021}
{Bonnaire} T.,  {Aghanim} N.,  {Kuruvilla} J.,   {Decelle} A.,  2021, arXiv
  e-prints, \href {https://ui.adsabs.harvard.edu/abs/2021arXiv211203926B} {p.
  arXiv:2112.03926}

\bibitem[\protect\citeauthoryear{{Colberg}}{{Colberg}}{2007}]{Colberg2007}
{Colberg} J.~M.,  2007, \mn@doi [\mnras] {10.1111/j.1365-2966.2006.11312.x},
  \href {http://adsabs.harvard.edu/abs/2007MNRAS.375..337C} {375, 337}

\bibitem[\protect\citeauthoryear{{Coles}, {Pearson}, {Borgani}, {Plionis}  \&
  {Moscardini}}{{Coles} et~al.}{1998}]{Coles1998}
{Coles} P.,  {Pearson} R.~C.,  {Borgani} S.,  {Plionis} M.,   {Moscardini} L.,
  1998, \mn@doi [\mnras] {10.1046/j.1365-8711.1998.01147.x}, \href
  {https://ui.adsabs.harvard.edu/abs/1998MNRAS.294..245C} {294, 245}

\bibitem[\protect\citeauthoryear{Cramer}{Cramer}{1946}]{Cramer1946}
Cramer H.,  1946, Princeton U. Press, Princeton, 500

\bibitem[\protect\citeauthoryear{{Fluri}, {Kacprzak}, {Refregier}, {Amara},
  {Lucchi}  \& {Hofmann}}{{Fluri} et~al.}{2018}]{machinelearning}
{Fluri} J.,  {Kacprzak} T.,  {Refregier} A.,  {Amara} A.,  {Lucchi} A.,
  {Hofmann} T.,  2018, \mn@doi [\prd] {10.1103/PhysRevD.98.123518}, \href
  {https://ui.adsabs.harvard.edu/abs/2018PhRvD..98l3518F} {98, 123518}

\bibitem[\protect\citeauthoryear{{Font-Ribera}, {McDonald}, {Mostek}, {Reid},
  {Seo}  \& {Slosar}}{{Font-Ribera} et~al.}{2014}]{Font2014}
{Font-Ribera} A.,  {McDonald} P.,  {Mostek} N.,  {Reid} B.~A.,  {Seo} H.-J.,
  {Slosar} A.,  2014, \mn@doi [\jcap] {10.1088/1475-7516/2014/05/023}, \href
  {http://adsabs.harvard.edu/abs/2014JCAP...05..023F} {5, 023}

\bibitem[\protect\citeauthoryear{{Foroozan}, {Krolewski}  \&
  {Percival}}{{Foroozan} et~al.}{2021}]{Foroozan2021}
{Foroozan} S.,  {Krolewski} A.,   {Percival} W.~J.,  2021, arXiv e-prints,
  \href {https://ui.adsabs.harvard.edu/abs/2021arXiv210611432F} {p.
  arXiv:2106.11432}

\bibitem[\protect\citeauthoryear{{Fukuda} et~al.,}{{Fukuda}
  et~al.}{1998}]{Kamiokande1998}
{Fukuda} Y.,  et~al., 1998, Physical Review Letters, \href
  {http://adsabs.harvard.edu/abs/1998PhRvL..81.1562F} {81, 1562}

\bibitem[\protect\citeauthoryear{{Gil-Mar{\'\i}n}, {Percival}, {Verde},
  {Brownstein}, {Chuang}, {Kitaura}, {Rodr{\'\i}guez-Torres}  \&
  {Olmstead}}{{Gil-Mar{\'\i}n} et~al.}{2017}]{bispectrum}
{Gil-Mar{\'\i}n} H.,  {Percival} W.~J.,  {Verde} L.,  {Brownstein} J.~R.,
  {Chuang} C.-H.,  {Kitaura} F.-S.,  {Rodr{\'\i}guez-Torres} S.~A.,
  {Olmstead} M.~D.,  2017, \mn@doi [\mnras] {10.1093/mnras/stw2679}, \href
  {https://ui.adsabs.harvard.edu/abs/2017MNRAS.465.1757G} {465, 1757}

\bibitem[\protect\citeauthoryear{{Gualdi}, {Gil-Mar{\'\i}n}, {Schuhmann},
  {Manera}, {Joachimi}  \& {Lahav}}{{Gualdi} et~al.}{2019}]{davideboss2018}
{Gualdi} D.,  {Gil-Mar{\'\i}n} H.,  {Schuhmann} R.~L.,  {Manera} M.,
  {Joachimi} B.,   {Lahav} O.,  2019, \mn@doi [\mnras] {10.1093/mnras/stz051},
  \href {https://ui.adsabs.harvard.edu/abs/2019MNRAS.484.3713G} {484, 3713}

\bibitem[\protect\citeauthoryear{{Hahn}, {Villaescusa-Navarro}, {Castorina}  \&
  {Scoccimarro}}{{Hahn} et~al.}{2020}]{Hahn2019}
{Hahn} C.,  {Villaescusa-Navarro} F.,  {Castorina} E.,   {Scoccimarro} R.,
  2020, \mn@doi [\jcap] {10.1088/1475-7516/2020/03/040}, \href
  {https://ui.adsabs.harvard.edu/abs/2020JCAP...03..040H} {2020, 040}

\bibitem[\protect\citeauthoryear{{Hartlap}, {Simon}  \& {Schneider}}{{Hartlap}
  et~al.}{2007}]{Hartlap2007}
{Hartlap} J.,  {Simon} P.,   {Schneider} P.,  2007, \mn@doi [\aap]
  {10.1051/0004-6361:20066170}, \href
  {https://ui.adsabs.harvard.edu/abs/2007A&A...464..399H} {464, 399}

\bibitem[\protect\citeauthoryear{{Hawken}, {Abdalla}, {H{\"u}tsi}  \&
  {Lahav}}{{Hawken} et~al.}{2012}]{Hawken2012}
{Hawken} A.~J.,  {Abdalla} F.~B.,  {H{\"u}tsi} G.,   {Lahav} O.,  2012, \mn@doi
  [\mnras] {10.1111/j.1365-2966.2012.20540.x}, \href
  {https://ui.adsabs.harvard.edu/abs/2012MNRAS.424....2H} {424, 2}

\bibitem[\protect\citeauthoryear{{Heavens}}{{Heavens}}{2009}]{Heavens2009}
{Heavens} A.,  2009, arXiv e-prints, \href
  {https://ui.adsabs.harvard.edu/abs/2009arXiv0906.0664H} {p. arXiv:0906.0664}

\bibitem[\protect\citeauthoryear{{Kaiser}}{{Kaiser}}{1987}]{Kaiser1987}
{Kaiser} N.,  1987, \mn@doi [\mnras] {10.1093/mnras/227.1.1}, \href
  {http://adsabs.harvard.edu/abs/1987MNRAS.227....1K} {227, 1}

\bibitem[\protect\citeauthoryear{Kaufman}{Kaufman}{1967}]{Kaufman1967}
Kaufman G.,  1967

\bibitem[\protect\citeauthoryear{{Krzewina} \& {Saslaw}}{{Krzewina} \&
  {Saslaw}}{1996}]{Krzewina1996}
{Krzewina} L.~G.,  {Saslaw} W.~C.,  1996, \mn@doi [\mnras]
  {10.1093/mnras/278.3.869}, \href
  {http://adsabs.harvard.edu/abs/1996MNRAS.278..869K} {278, 869}

\bibitem[\protect\citeauthoryear{{Libeskind} et~al.,}{{Libeskind}
  et~al.}{2018}]{Libeskind2018}
{Libeskind} N.~I.,  et~al., 2018, \mn@doi [\mnras] {10.1093/mnras/stx1976},
  \href {http://adsabs.harvard.edu/abs/2018MNRAS.473.1195L} {473, 1195}

\bibitem[\protect\citeauthoryear{{Massara}, {Villaescusa-Navarro}, {Ho},
  {Dalal}  \& {Spergel}}{{Massara} et~al.}{2021}]{Massara2020}
{Massara} E.,  {Villaescusa-Navarro} F.,  {Ho} S.,  {Dalal} N.,   {Spergel}
  D.~N.,  2021, \mn@doi [\prl] {10.1103/PhysRevLett.126.011301}, \href
  {https://ui.adsabs.harvard.edu/abs/2021PhRvL.126a1301M} {126, 011301}

\bibitem[\protect\citeauthoryear{{Naidoo}}{{Naidoo}}{2019}]{mistree}
{Naidoo} K.,  2019, \mn@doi [The Journal of Open Source Software]
  {10.21105/joss.01721}, \href
  {https://ui.adsabs.harvard.edu/abs/2019JOSS....4.1721N} {4, 1721}

\bibitem[\protect\citeauthoryear{{Naidoo}, {Whiteway}, {Massara}, {Gualdi},
  {Lahav}, {Viel}, {Gil-Mar{\'\i}n}  \& {Font-Ribera}}{{Naidoo}
  et~al.}{2020}]{Naidoo2020}
{Naidoo} K.,  {Whiteway} L.,  {Massara} E.,  {Gualdi} D.,  {Lahav} O.,  {Viel}
  M.,  {Gil-Mar{\'\i}n} H.,   {Font-Ribera} A.,  2020, \mn@doi [\mnras]
  {10.1093/mnras/stz3075}, \href
  {https://ui.adsabs.harvard.edu/abs/2020MNRAS.491.1709N} {491, 1709}

\bibitem[\protect\citeauthoryear{{Petri}, {Haiman}, {Hui}, {May}  \&
  {Kratochvil}}{{Petri} et~al.}{2013}]{minkowski}
{Petri} A.,  {Haiman} Z.,  {Hui} L.,  {May} M.,   {Kratochvil} J.~M.,  2013,
  \mn@doi [\prd] {10.1103/PhysRevD.88.123002}, \href
  {https://ui.adsabs.harvard.edu/abs/2013PhRvD..88l3002P} {88, 123002}

\bibitem[\protect\citeauthoryear{{Planck Collaboration} et~al.,}{{Planck
  Collaboration} et~al.}{2020}]{PlanckPara2018}
{Planck Collaboration} et~al., 2020, \mn@doi [\aap]
  {10.1051/0004-6361/201833910}, \href
  {https://ui.adsabs.harvard.edu/abs/2020A&A...641A...6P} {641, A6}

\bibitem[\protect\citeauthoryear{Rao}{Rao}{1945}]{Rao1945}
Rao C.~R.,  1945, Reson. J. Sci. Educ, 20, 78

\bibitem[\protect\citeauthoryear{{Tegmark}, {Taylor}  \& {Heavens}}{{Tegmark}
  et~al.}{1997}]{Tegmark1997}
{Tegmark} M.,  {Taylor} A.~N.,   {Heavens} A.~F.,  1997, \mn@doi [\apj]
  {10.1086/303939}, \href
  {https://ui.adsabs.harvard.edu/abs/1997ApJ...480...22T} {480, 22}

\bibitem[\protect\citeauthoryear{{Ueda} \& {Itoh}}{{Ueda} \&
  {Itoh}}{1997}]{Ueda1997}
{Ueda} H.,  {Itoh} M.,  1997, \mn@doi [\pasj] {10.1093/pasj/49.2.131}, \href
  {http://adsabs.harvard.edu/abs/1997PASJ...49..131U} {49, 131}

\bibitem[\protect\citeauthoryear{{Uhlemann}, {Friedrich},
  {Villaescusa-Navarro}, {Banerjee}  \& {Codis}}{{Uhlemann}
  et~al.}{2020}]{Uhlemann2019}
{Uhlemann} C.,  {Friedrich} O.,  {Villaescusa-Navarro} F.,  {Banerjee} A.,
  {Codis} S.,  2020, \mn@doi [\mnras] {10.1093/mnras/staa1155}, \href
  {https://ui.adsabs.harvard.edu/abs/2020MNRAS.495.4006U} {495, 4006}

\bibitem[\protect\citeauthoryear{{Villaescusa-Navarro}
  et~al.,}{{Villaescusa-Navarro} et~al.}{2020}]{Quijote2019}
{Villaescusa-Navarro} F.,  et~al., 2020, \mn@doi [\apjs]
  {10.3847/1538-4365/ab9d82}, \href
  {https://ui.adsabs.harvard.edu/abs/2020ApJS..250....2V} {250, 2}

\bibitem[\protect\citeauthoryear{{van de Weygaert}, {Jones}  \&
  {Mart{\'\i}nez}}{{van de Weygaert} et~al.}{1992}]{Weygaert1992}
{van de Weygaert} R.,  {Jones} B. J.~T.,   {Mart{\'\i}nez} V.~J.,  1992,
  \mn@doi [Physics Letters A] {10.1016/0375-9601(92)90584-9}, \href
  {https://ui.adsabs.harvard.edu/abs/1992PhLA..169..145V} {169, 145}

\makeatother
\end{thebibliography}

%%%%%%%%%%%%%%%%%%%%%%%%%%%%%%%%%%%%%%%%%%%%%%%%%%%
%%%%%%%%%%%%%%%%%%%  APPENDICES %%%%%%%%%%%%%%%%%%%%%%
\appendix

\section{Masking Small Scales Using Groups}
\label{appendix:small_scales}

Cosmological simulations tend to suffer from inaccuracies at small scales due to limitations in resolution and complex baryonic effects which require expensive hydrodynamics simulations. For these reasons it is important to be able to remove these troublesome scales from our analysis. For traditional $N$-point statistics this is relatively straightforward as this simply requires placing scale cuts on the data vector. However the MST is different as there is no way to mask small scales once the MST has been constructed; instead, the input data vector needs to be prepared such that scales are not present when the MST is constructed. To remove these scales from our analysis we implement a grouping scheme where points separated by less than $l_{\min}$ are combined.

In this section we investigate the effectiveness of masking small scales by grouping using two Levy-Flight simulations as a proxy for $N$-body simulations with accurate and inaccurate small scale effects (such as baryonic effects). These simulations were chosen as their clustering properties are relatively simple to control and are fast to generate. We generate 50 realisations of two Levy Flight simulation models: (1) the standard Levy Flight (LF) and (2) the Adjusted Levy Flight \citep[ALF; see][for more details]{Naidoo2020}. The simulations are generated using \textsc{MiSTree} \citep{mistree}. For each simulation 50,000 points are generated in a box of length 75 with parameters $t_{0}=0.24$ and $\alpha=1.6$ for the LF simulations and parameters $t_{0}=0.325$, $t_{s}=0.015$, $\alpha=1.5$, $\beta=0.45$, and $\gamma=1.3$ for the ALF.

\begin{figure*}
	\centering
	\includegraphics[width=\textwidth]{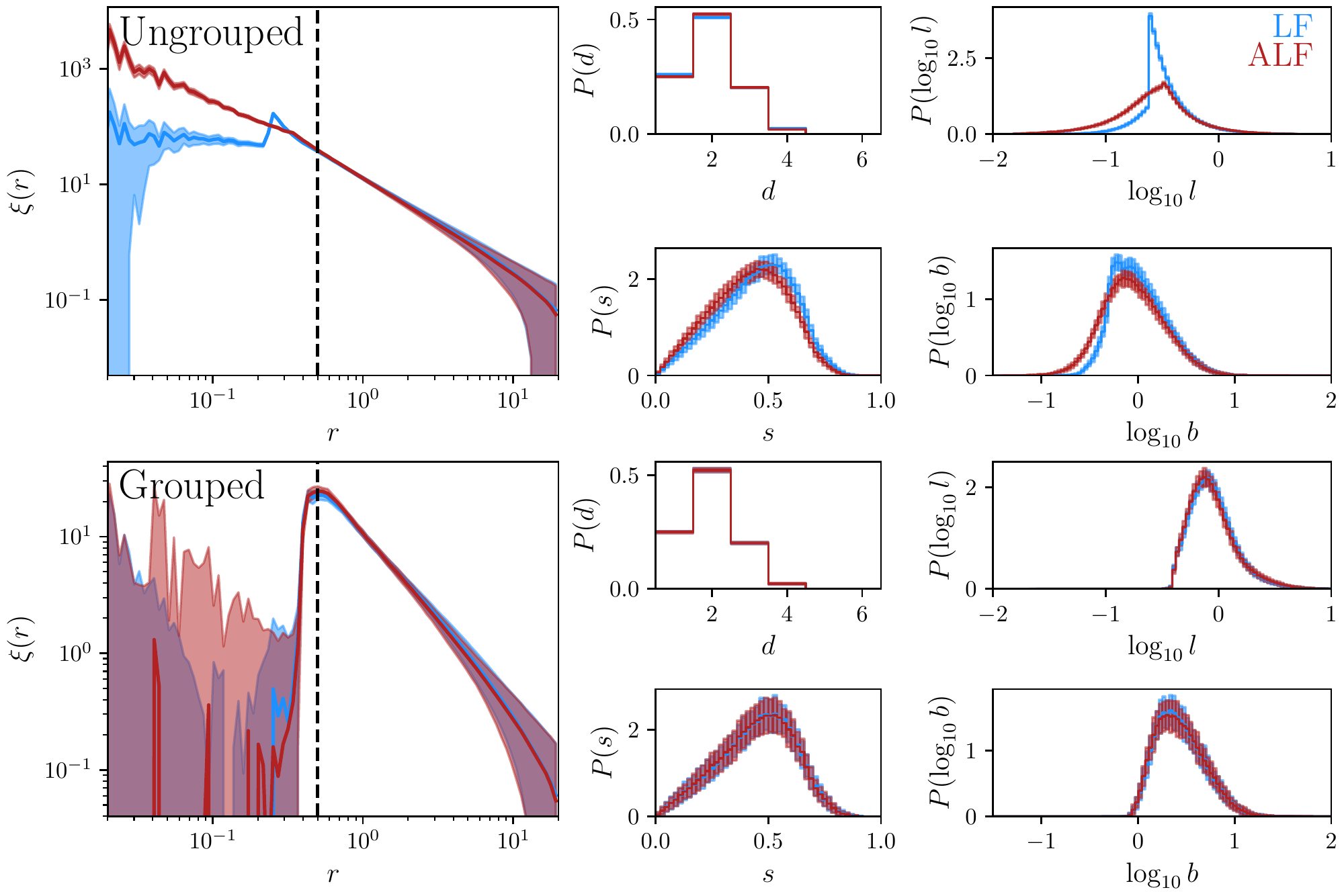}
	\caption{The effect of grouping Levy Flight simulations are shown for the 2-point correlation function (2PCF) and MST statistics. Two sets of Levy Flight simulations are used: the standard Levy Flight (LF) simulations are shown in blue and the adjusted Levy Flight (ALF) simulations are shown in red. They exhibit almost identical large scale 2PCF $\xi(r)$ by design but have very different small scale correlation properties. The differences between the ungrouped simulations are shown in the top subpanels. On the left is the 2PCF and on the right are the four MST statistics degree ($d$), edge length ($l$), branch shape ($s$), and branch length ($b$). The differences between the grouped simulations are shown in the bottom subpanels. On the left is the 2PCF and on the right are the four MST statistics $d$, $l$, $s$, and $b$. These plots show that grouping provides an effective solution for masking small scales as the small scale differences between the simulations do not appear once grouping has been performed.}
	\label{fig:small_scales}
\end{figure*}

The 2-point correlation function (2PCF) and MST statistics are measured for both simulations and compared in Fig. \ref{fig:small_scales}. The simulations are designed to exhibit very different small scale properties but identical large scale correlation functions.

All points in the simulations with separations less than $0.4$ are grouped and replaced by a single point with the average position of the group members. A new catalogue of points is now constructed with these grouped points and any remaining ungrouped points. For both simulations this results in a catalogue of roughly $\sim 13,000$ points. To ensure differences between the statistics are not due to slight differences in density we sample at random 12,000 points from each catalogue and measure the 2PCF and MST statistics. In Fig \ref{fig:small_scales} the statistics before and after grouping are shown. Prior to grouping the simulations exhibit different small scale clustering properties, which is seen by differences in the 2PCF and MST statistics, while after grouping all the statistics are consistent with each other. This shows that grouping effectively masks the small scale differences between the simulations.

\section{Minimum Spanning Tree Partial Derivatives}
\label{appendix:partial_derivatives}

\begin{figure*}
	\centering
	\includegraphics[width=\textwidth]{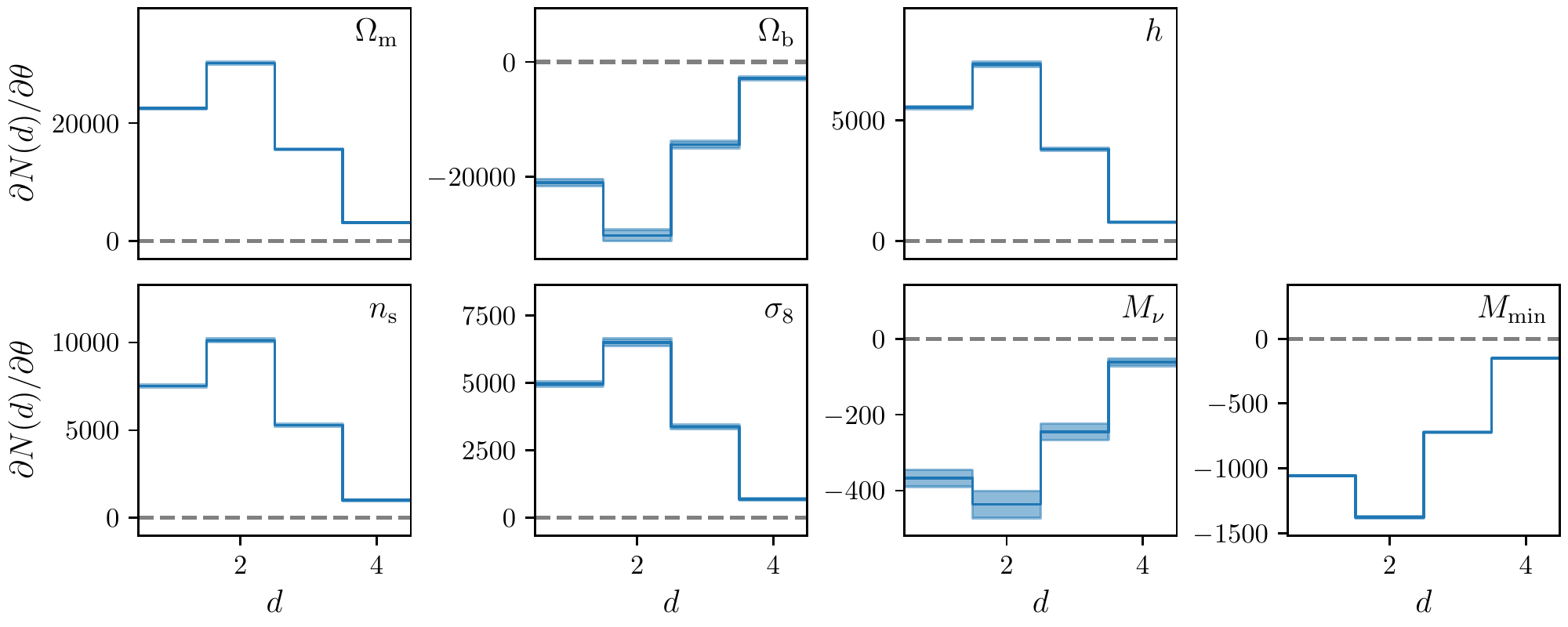}
	\caption{Derivatives for the MST degree $d$ are shown for the six $\nu\Lambda$CDM parameters ($h$, $n_{s}$, $\Omega_{\rm b}$, $\Omega_{\rm m}$, $\sigma_{8}$, and $M_{\nu}$) in redshift space and an additional nuisance parameter $M_{\min}$. The sensitivity to each parameter can be assessed by the significance of deviations away from $\partial N(d)/\partial\theta=0$ (dotted black line).}
	\label{mst_d_RSD_derivatives}
\end{figure*}

\begin{figure*}
	\centering
	\includegraphics[width=\textwidth]{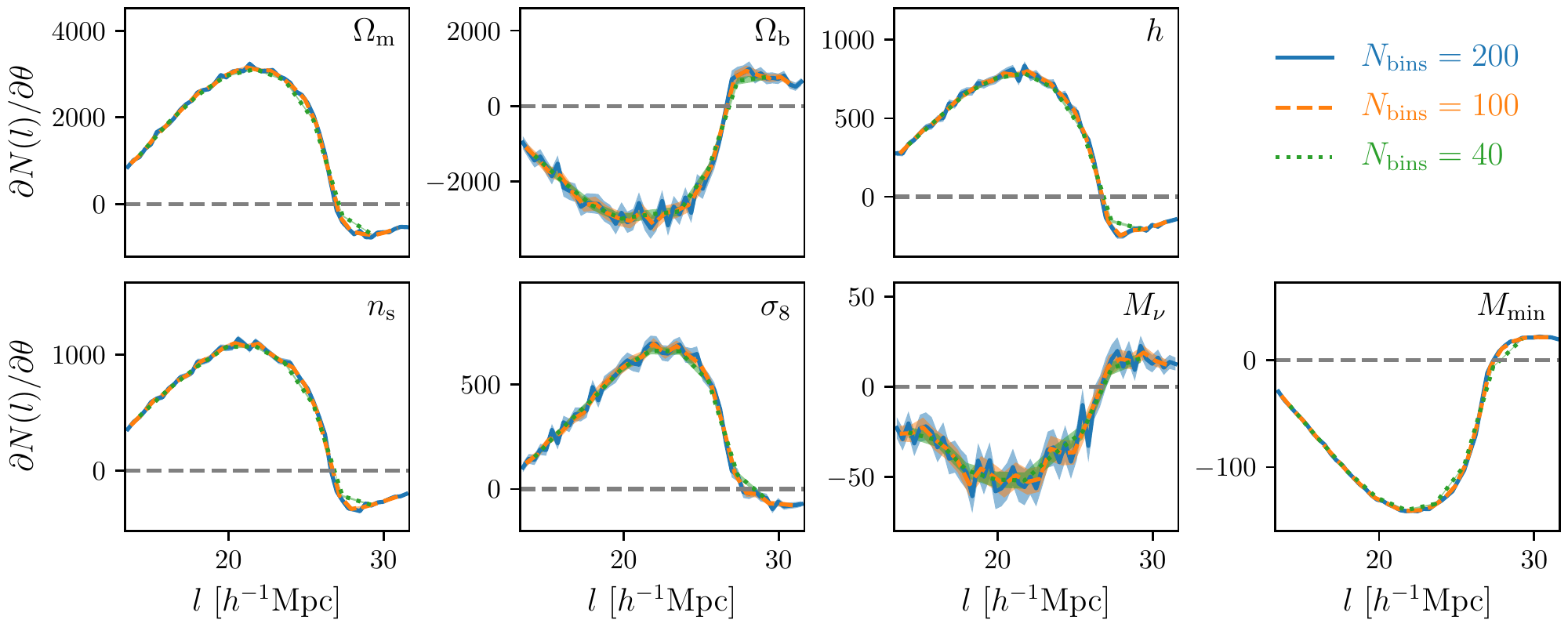}
	\caption{Derivatives for the MST edge length $l$ are shown for the six $\nu\Lambda$CDM parameters ($h$, $n_{s}$, $\Omega_{\rm b}$, $\Omega_{\rm m}$, $\sigma_{8}$, and $M_{\nu}$) in redshift space and an additional nuisance parameter $M_{\min}$. The sensitivity to each parameter can be assessed by the significance of deviations away from $\partial N(l)/\partial\theta=0$ (dotted black line). The derivatives are shown for three binning schemes ($N_{\rm bins}$): 200 (blue), 100 (orange), and 40 (green).}
	\label{mst_l_RSD_derivatives}
\end{figure*}

\begin{figure*}
	\centering
	\includegraphics[width=\textwidth]{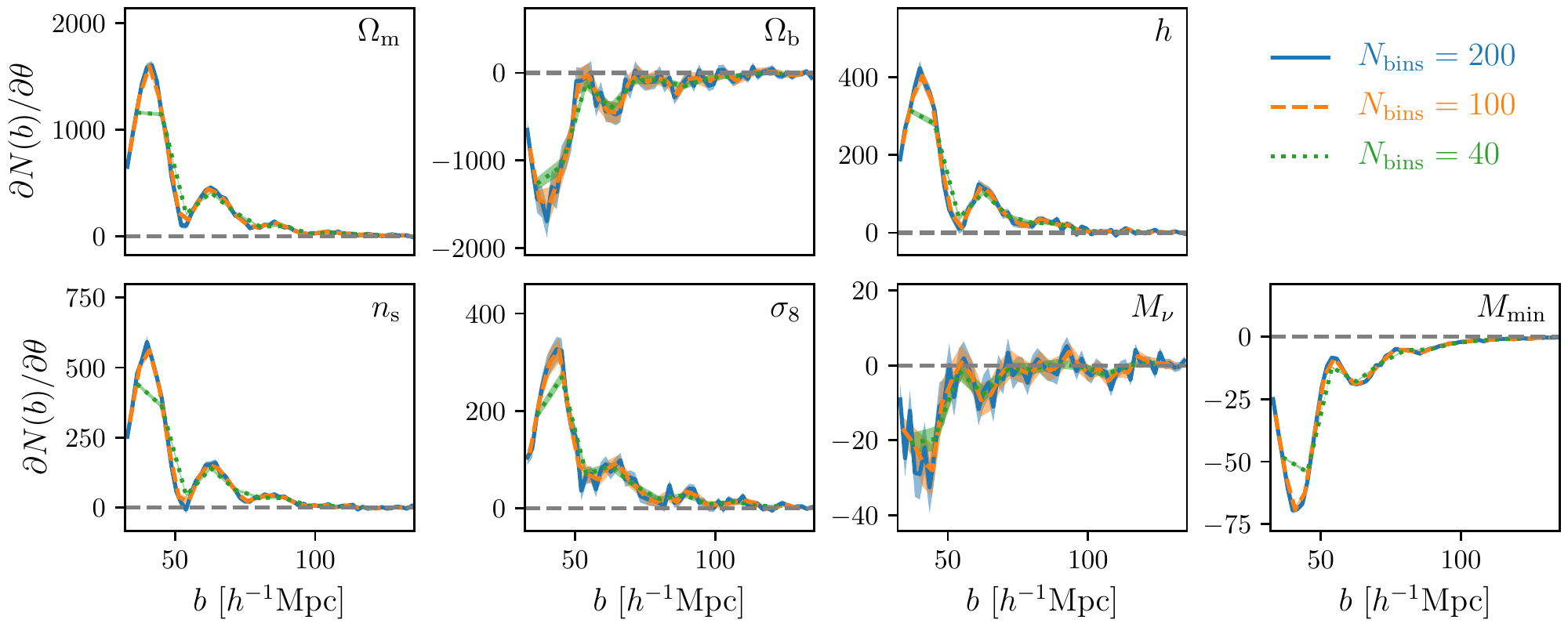}
	\caption{Derivatives for the MST branch length $b$ are shown for the six $\nu\Lambda$CDM parameters ($h$, $n_{s}$, $\Omega_{\rm b}$, $\Omega_{\rm m}$, $\sigma_{8}$, and $M_{\nu}$) in redshift space and an additional nuisance parameter $M_{\min}$. The sensitivity to each parameter can be assessed by the significance of deviations away from $\partial N(b)/\partial\theta=0$ (dotted black line). The derivatives are shown for three binning schemes ($N_{\rm bins}$): 200 (blue), 100 (orange), and 40 (green).}
	\label{mst_b_RSD_derivatives}
\end{figure*}

\begin{figure*}
	\centering
	\includegraphics[width=\textwidth]{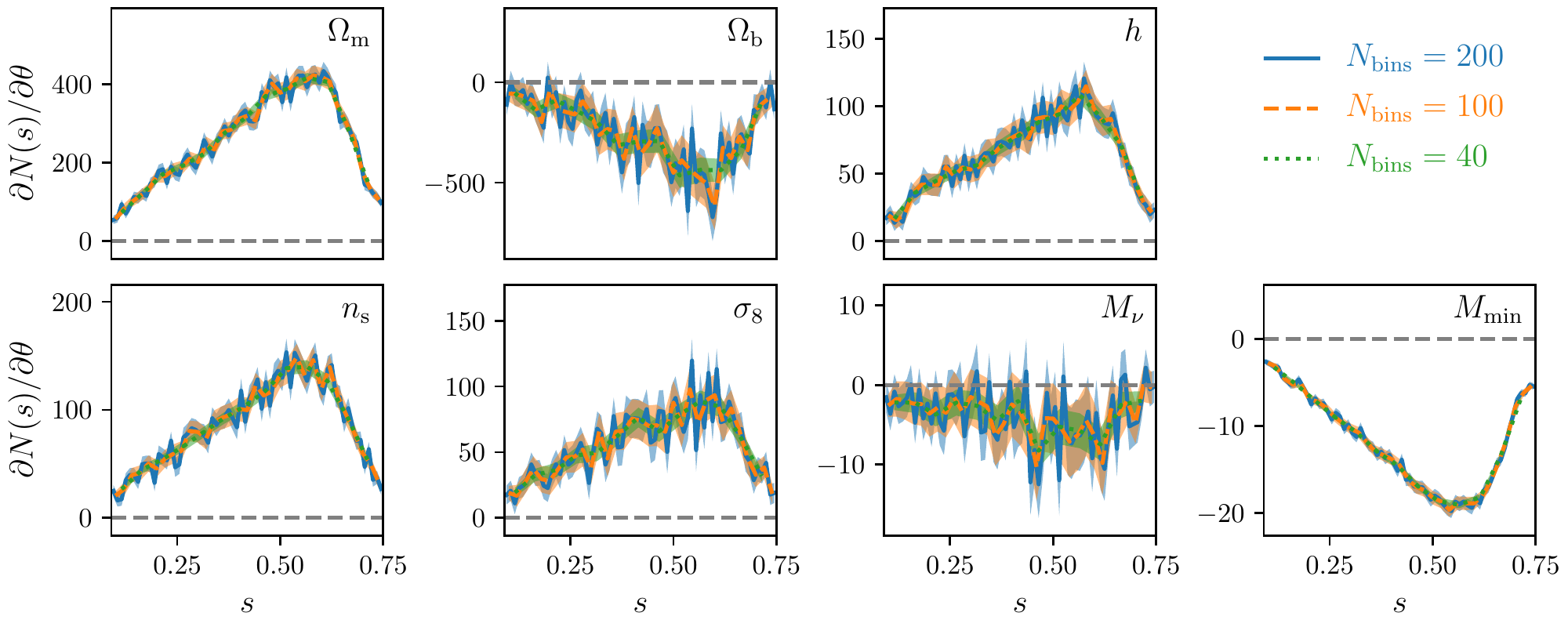}
	\caption{Derivatives for the MST branch shape $s$ are shown for the six $\nu\Lambda$CDM parameters ($h$, $n_{s}$, $\Omega_{\rm b}$, $\Omega_{\rm m}$, $\sigma_{8}$, and $M_{\nu}$) in redshift space and an additional nuisance parameter $M_{\min}$. The sensitivity to each parameter can be assessed by the significance of deviations away from $\partial N(s)/\partial\theta=0$ (dotted black line). The derivatives are shown for three binning schemes ($N_{\rm bins}$): 200 (blue), 100 (orange), and 40 (green).}
	\label{mst_s_RSD_derivatives}
\end{figure*}

The partial derivative estimates for the MST statistics are shown for the distribution of degree $d$ in Fig. \ref{mst_d_RSD_derivatives}, for the distribution of edges $l$ in Fig. \ref{mst_l_RSD_derivatives}, for the distribution of branches $b$ in Fig. \ref{mst_b_RSD_derivatives}, and for the distribution of branch shapes $s$ in Fig. \ref{mst_s_RSD_derivatives}. The derivatives for neutrino mass are shown for estimator 2 (Eq. \ref{eq:derivative_eq2}) with $M_{\nu}=0.2\, {\rm eV}$.

For a fixed MST statistic, the partial derivatives of that statistic (with respect to the various cosmological parameters) all have similar shapes. The similarity can be explained if we think about the construction of the MST as an optimisation problem: differences in parameters will lead to trees with longer (poorly optimised) or shorter edges (highly optimised). Depending on the outcome the MST statistics will to first order be pulled to smaller or larger values but since there are roughly the same number of points this change has to be counter balanced with a reduction in the opposite direction.
Although the profiles are similar in their general shape if we instead look to the relative peaks and troughs with respect to zero we can see that each parameter behaves slightly differently; take for example the partial derivatives for edge length with respect to $\Omega_{\rm m}$ and $M_{\min}$ where the troughs for $M_{\min}$ is much deeper. The similarities suggest that the MST statistics data vectors could be compressed into a few values (this possibility will be investigated in future work). In any case the similarities are not a cause for concern as the interdependencies of these parameters are taken into account in the Fisher matrix calculation and would appear as covariances.

\begin{figure}
  \includegraphics[width=\columnwidth]{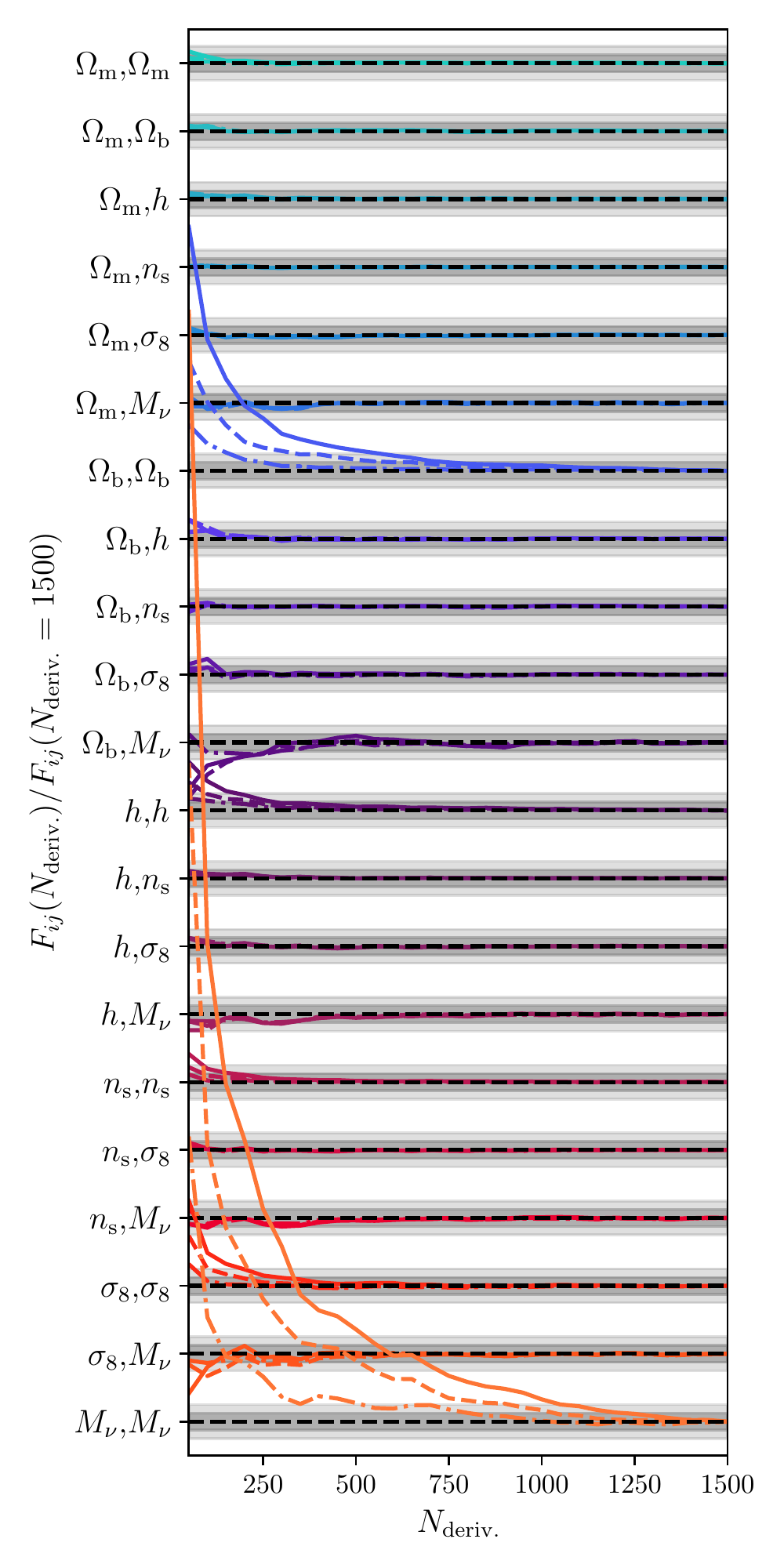}
  \caption{Convergence of the MST Fisher matrix elements as a function of the number of data vectors used for the derivative estimates. The x-axis shows the number of data vectors used to compute the partial derivatives ($N_{\rm deriv.}$) of the MST statistics at redshift $z=0.5$ (similar results are obtained for $z=0$ and $z=1$). On the y-axis the components of the Fisher matrix are shown. The grey bands show $10\%$ (light) and $5\%$ (dark) convergence from the Fisher matrix values after all 1500 derivative estimates are used. The full lines show the convergence for MST statistics with $N_{\rm bins}=200$, dashed lines with $N_{\rm bins}=100$ and dashed-dotted lines with $N_{\rm bins}=40$. For most of these statistics convergence is reached relatively quickly with the exception of $\Omega_{\rm b}$ and $M_{\nu}$ whose effects are generally more subtle. In both cases we show convergence within $5\%$ is achieved for $N_{\rm deriv.}>1300$.}
  \label{mst_derivative_convergence}
\end{figure}

To test whether the partial derivatives have converged we compute the Fisher matrix from a subset of the total simulations available. In Fig.~\ref{mst_derivative_convergence} we show the convergence of the MST partial derivatices by showing whether the individual components of the Fisher matrix are converging to their final values. Convergence is loosely defined to be Fisher matrix components which have settled to within $5\%$ of their final value when $N_{\rm deriv.}$ (the number of data vectors used to compute partial derivatives) is greater than $1300$. We find that estimator 2 (Eq. \ref{eq:derivative_eq2}) with $M_{\nu}=0.2\, {\rm eV}$ provides the best convergence properties at all redshifts. Estimator 2 (Eq. \ref{eq:derivative_eq2}) with $M_{\nu}=0.1\, {\rm eV}$ and Estimator 3 (Eq. \ref{eq:derivative_eq3}) while being more accurate estimators tend to show poorer convergence at redshift $z=0$ where the constraining power for $M_{\nu}$ is generally worse. This is likely the combination of two factors: (1) at low redshifts the halo catalogues are larger meaning more groupings take place resulting in some loss of information due to percolation and (2) gravity infall scrambles the effect of neutrinos free streaming making their effect harder to detect. We believe the former is the more important feature but this will need to be investigated in later work to understand how to limit the effects of percolation on the MST, especially for dealing with scenarios where $l_{\min}$ is larger than the mean separation between haloes.

\section{Fisher Matrix Constraints from the MST}

Additional constraints from the MST and power spectrum multipoles are presented in this section. In addition to the measurements made in the paper at redshift $z=0.5$, we also make measurements at redshift $z=0$ and $z=1$, and finally combine the analysis over the three redshifts.

In Tab. \ref{mv_estimator_constraints} we compare the constraints of the MST and power spectrum multipoles for the different neutrino mass derivative estimates presented in this paper.

In Tab. \ref{tab:summarise_constraints} we compare the constraints of the MST and power spectrum multipoles for the six $\nu\Lambda$CDM parameters and the nuisance minimum halo mass parameter $M_{\min}$.

\begin{table*}
	\centering
	\caption{Constraints on $M_{\nu}$ using different estimators for $\partial\datavector{}/\partial M_{\nu}$ at $z=0$: estimator 1 (Eq. \ref{eq:derivative_eq1}) using $M_{\nu}=0.1$, $0.2$, and $0.4\,{\rm eV}$, estimator 2 (Eq. \ref{eq:derivative_eq2}) using $M_{\nu}=0.1$ and $0.2\,{\rm eV}$, and estimator 3 (Eq. \ref{eq:derivative_eq3}) using $M_{\nu}=0.1\,{\rm eV}$. Typically estimator 3 has been used in previous studies but in this paper we use estimator 2 with $M_{\nu}=0.2\,{\rm eV}$ since the accuracy of this estimator is consistent with Eq. \ref{eq:derivative_eq1} used for the other parameters. All the estimators in redshift space show tighter constraints for the MST than the power spectrum. Furthermore, when they are combined the constraints appear to be dominated by the MST.}
	\label{mv_estimator_constraints}
	\begin{tabular}{ll|ccc|cc|c}
		\hline
		\Tstrut \multirow{2}{*}{Statistics} & \multirow{2}{*}{Redshift}  & \multicolumn{3}{c|}{Est. 1 (Eq. \ref{eq:derivative_eq1})} &  \multicolumn{2}{c|}{Est. 2 (Eq. \ref{eq:derivative_eq2})}  &  \multicolumn{1}{c}{Est. 3 (Eq. \ref{eq:derivative_eq3})}\\
		\Bstrut & & $0.1\,{\rm eV}$ & $0.2\,{\rm eV}$ & $0.4\,{\rm eV}$ & $0.1\,{\rm eV}$ & $0.2\,{\rm eV}$ & $0.1\,{\rm eV}$ \\
		\hline
		\Tstrut ${\rm MST}(l)$ & 0.5 & 0.5 & 1.0 & 1.8 & 0.29 & 0.6 & 0.22 \\
    ${\rm MST}(b)$ & 0.5 & 0.45 & 0.7 & 1.4 & 0.26 & 0.38 & 0.2 \\
    ${\rm MST}(l, b)$ & 0.5 & 0.33 & 0.56 & 1.1 & 0.19 & 0.31 & 0.15 \\
    ${\rm MST}(l, b, s)$ & 0.5 & 0.23 & 0.46 & 0.79 & 0.13 & 0.25 & 0.099 \\
    ${\rm MST}(d, l, b, s)$ & 0.5 & 0.23 & 0.45 & 0.76 & 0.13 & 0.25 & 0.099 \\
    $P_{\ell=0,2}(k)$ & 0.5 & 0.82 & 1.6 & 1.4 & 0.44 & 1.0 & 0.33 \\
    \Bstrut $P_{\ell=0,2}(k)+{\rm MST}(d,l,b,s)$ & 0.5 & 0.21 & 0.38 & 0.44 & 0.12 & 0.23 & 0.093 \\
		\hline
	\end{tabular}
\end{table*}

\begin{table*}
	\centering
	\caption{Separate and combined constraints for parameters from the $\nu\Lambda$CDM model determined from measurements of the power spectrum (multipoles in redshift space) and MST at redshift $z=0$, $0.5$, and $1$. The constraints are obtained using Eq. \ref{eq:derivative_eq0} for all of the parameters except $M_{\nu}$ which are obtained using Eq.  \ref{eq:derivative_eq2} with $M_{\nu}=0.2\, {\rm eV}$. For the standard $\Lambda$CDM parameters we obtain competitive constraints from the MST at all redshifts, with the exception of $\Omega_{\rm m}$ and $\sigma_{8}$ where the power spectrum dominates; however for $M_{\nu}$ the MST dominates. When measurements from the different redshifts are combined we find the MST is competitive for all parameters, including $\Omega_{\rm m}$ and $\sigma_{8}$, but still dominates constraints on $M_{\nu}$.}
	\label{tab:summarise_constraints}
	\begin{tabular}{llcccccccc}
		\hline
		\TBstrut Statistics & Redshift & $N_{\rm bins}$ & $\Delta \Omega_{\rm m}$ & $\Delta \Omega_{\rm b}$ & $\Delta h$ & $\Delta n_{\rm s}$ & $\Delta \sigma_{8}$ & $\Delta M_{\nu}$  $[\rm eV]$ & $\Delta M_{\min}$ \\
		\hline
    \Tstrut ${\rm MST}(l)$ & 0 & 1 & 0.076 & 0.018 & 0.13 & 0.12 & 0.14 & 0.53 & 1.2 \\
    ${\rm MST}(b)$ & 0 & 1 & 0.062 & 0.015 & 0.14 & 0.14 & 0.098 & 0.45 & 1.4 \\
    ${\rm MST}(l, b)$ & 0 & 1 & 0.048 & 0.011 & 0.089 & 0.091 & 0.079 & 0.33 & 0.75 \\
    ${\rm MST}(l, b, s)$ & 0 & 1 & 0.036 & 0.0083 & 0.073 & 0.067 & 0.067 & 0.24 & 0.59 \\
    ${\rm MST}(d, l, b, s)$ & 0 & 1 & 0.036 & 0.0083 & 0.073 & 0.065 & 0.067 & 0.23 & 0.55 \\
    $P_{\ell=0,2}(k)$ & 0 & & 0.029 & 0.013 & 0.13 & 0.079 & 0.061 & 1.1 & 0.22 \\
    $P_{\ell=0,2}(k)+{\rm MST}(d,l,b,s)$ & 0 & 1 & 0.015 & 0.0059 & 0.054 & 0.036 & 0.016 & 0.22 & 0.14 \\
    $P_{\ell=0,2}(k)+{\rm MST}(d,l,b,s)$ & 0 & 2 & 0.016 & 0.0074 & 0.069 & 0.041 & 0.019 & 0.29 & 0.14 \\
    $P_{\ell=0,2}(k)+{\rm MST}(d,l,b,s)$ & 0 & 5 & 0.017 & 0.0099 & 0.093 & 0.046 & 0.023 & 0.37 & 0.15 \\
    ${\rm MST}(l)$ & 0.5 & 1 & 0.062 & 0.024 & 0.19 & 0.15 & 0.14 & 0.6 & 1.2 \\
    ${\rm MST}(b)$ & 0.5 & 1 & 0.065 & 0.016 & 0.15 & 0.12 & 0.099 & 0.38 & 1.2 \\
    ${\rm MST}(l, b)$ & 0.5 & 1 & 0.042 & 0.013 & 0.11 & 0.091 & 0.079 & 0.31 & 0.73 \\
    ${\rm MST}(l, b, s)$ & 0.5 & 1 & 0.035 & 0.0092 & 0.089 & 0.068 & 0.062 & 0.25 & 0.59 \\
    ${\rm MST}(d, l, b, s)$ & 0.5 & 1 & 0.035 & 0.0092 & 0.088 & 0.067 & 0.061 & 0.25 & 0.58 \\
    $P_{\ell=0,2}(k)$ & 0.5 & & 0.023 & 0.014 & 0.14 & 0.09 & 0.065 & 1.0 & 0.2 \\
    $P_{\ell=0,2}(k)+{\rm MST}(d,l,b,s)$ & 0.5 & 1 & 0.013 & 0.0068 & 0.066 & 0.043 & 0.017 & 0.23 & 0.13 \\
    $P_{\ell=0,2}(k)+{\rm MST}(d,l,b,s)$ & 0.5 & 2 & 0.014 & 0.0086 & 0.084 & 0.051 & 0.022 & 0.33 & 0.14 \\
    $P_{\ell=0,2}(k)+{\rm MST}(d,l,b,s)$ & 0.5 & 5 & 0.016 & 0.01 & 0.1 & 0.058 & 0.03 & 0.46 & 0.14 \\
    ${\rm MST}(l)$ & 1 & 1 & 0.055 & 0.018 & 0.17 & 0.12 & 0.097 & 0.52 & 1.2 \\
    ${\rm MST}(b)$ & 1 & 1 & 0.066 & 0.016 & 0.16 & 0.11 & 0.089 & 0.38 & 1.1 \\
    ${\rm MST}(l, b)$ & 1 & 1 & 0.04 & 0.011 & 0.11 & 0.079 & 0.063 & 0.3 & 0.77 \\
    ${\rm MST}(l, b, s)$ & 1 & 1 & 0.034 & 0.0088 & 0.086 & 0.064 & 0.051 & 0.24 & 0.62 \\
    ${\rm MST}(d, l, b, s)$ & 1 & 1 & 0.034 & 0.0087 & 0.08 & 0.063 & 0.051 & 0.24 & 0.58 \\
    $P_{\ell=0,2}(k)$ & 1 & & 0.024 & 0.016 & 0.18 & 0.14 & 0.065 & 0.88 & 0.25 \\
    $P_{\ell=0,2}(k)+{\rm MST}(d,l,b,s)$ & 1 & 1 & 0.014 & 0.0063 & 0.057 & 0.046 & 0.02 & 0.23 & 0.12 \\
    $P_{\ell=0,2}(k)+{\rm MST}(d,l,b,s)$ & 1 & 2 & 0.016 & 0.0078 & 0.073 & 0.054 & 0.025 & 0.32 & 0.13 \\
    $P_{\ell=0,2}(k)+{\rm MST}(d,l,b,s)$ & 1 & 5 & 0.017 & 0.01 & 0.098 & 0.069 & 0.033 & 0.43 & 0.14 \\
    ${\rm MST}(l)$ & all & 1 & 0.034 & 0.01 & 0.083 & 0.064 & 0.035 & 0.29 & 0.46 \\
    ${\rm MST}(b)$ & all & 1 & 0.031 & 0.0083 & 0.075 & 0.061 & 0.033 & 0.21 & 0.45 \\
    ${\rm MST}(l, b)$ & all & 1 & 0.023 & 0.0065 & 0.055 & 0.045 & 0.022 & 0.17 & 0.32 \\
    ${\rm MST}(l, b, s)$ & all & 1 & 0.019 & 0.0049 & 0.043 & 0.035 & 0.019 & 0.14 & 0.26 \\
    ${\rm MST}(d, l, b, s)$ & all & 1 & 0.019 & 0.0049 & 0.043 & 0.035 & 0.019 & 0.14 & 0.26 \\
    $P_{\ell=0,2}(k)$ & all & & 0.014 & 0.0083 & 0.07 & 0.048 & 0.026 & 0.37 & 0.15 \\
    $P_{\ell=0,2}(k)+{\rm MST}(d,l,b,s)$ & all & 1 & 0.0081 & 0.0039 & 0.033 & 0.025 & 0.01 & 0.13 & 0.071 \\
    $P_{\ell=0,2}(k)+{\rm MST}(d,l,b,s)$ & all & 2 & 0.0091 & 0.0049 & 0.041 & 0.029 & 0.013 & 0.17 & 0.073 \\
    \Bstrut $P_{\ell=0,2}(k)+{\rm MST}(d,l,b,s)$ & all & 5 & 0.01 & 0.0062 & 0.052 & 0.033 & 0.015 & 0.21 & 0.075 \\
		\hline
	\end{tabular}
\end{table*}

%%%%%%%%%%%%%%%%%%%%%%%%%%%%%%%%%%%%%%%%%%%%%%%%%%

% Don't change these lines
\bsp	% typesetting comment
\label{lastpage}
\end{document}